\documentclass[reprint,aps,prc,twocolumn,floatfix,showpacs,preprintnumbers,amsmath,amssymb,nofootinbib,groupedaddress]{revtex4-1}

\usepackage{color}
\usepackage{graphicx}
\usepackage{dcolumn}    
\usepackage{multirow}
\usepackage{bm}         
\usepackage{sidecap}
\usepackage{hyperref}   
\usepackage{xcolor}
\usepackage{amsmath}
\usepackage{color,colortbl}
\usepackage{booktabs} 
\usepackage{hhline}
\usepackage{ulem}
\usepackage{hyperref}

\definecolor{dark-red}{rgb}{0.,0.,0}
\definecolor{dark-blue}{rgb}{0.,0.,1}
\definecolor{medium-blue}{rgb}{0,0,1}
\definecolor{gray}{rgb}{0.85,0.85,0.85}

\hypersetup{
    colorlinks, linkcolor={dark-red},
    citecolor={dark-blue}, urlcolor={medium-blue}
}

\begin{document}

\title{Gamow-Teller excitations at finite temperature: Competition between pairing and temperature effects}

\author{E. Y{\"{u}}ksel}
\email{eyuksel@yildiz.edu.tr}
\affiliation {Yildiz Technical University, Faculty of Arts and Science, Department of Physics, Davutpasa Campus, TR-34220, Esenler/Istanbul, Turkey}

\author{N. Paar}
\affiliation {Department of Physics, Faculty of Science, University of Zagreb, Bijeni\v{c}ka c. 32, 10000 Zagreb, Croatia}

\author{G. Col\`o }
\affiliation {Dipartimento di Fisica, Universit\`a degli 
             Studi di Milano, via Celoria 16, I-20133 Milano, Italy, \\ INFN Sezione di Milano, Via Celoria 16, 20133 Milano, Italy}

\author{E. Khan}
\affiliation {Institut de Physique Nucl\'eaire, Universit\'e Paris-Sud, IN2P3-CNRS, \\ Universite Paris-Saclay, F-91406 Orsay Cedex, France}

\author{Y. F. Niu}
\affiliation {School of Nuclear Science and Technology, Lanzhou University, Lanzhou 730000, China, \\
ELI-NP, “Horia Hulubei” National Institute for Physics and Nuclear Engineering,
30 Reactorului Street, RO-077125, Bucharest-Magurele, Romania}

\date{\today}

\begin{abstract}
The relativistic and nonrelativistic finite temperature proton-neutron quasiparticle random phase approximation (FT-PNQRPA) methods are developed to study the interplay of the pairing and temperature effects on the Gamow-Teller excitations in open-shell nuclei, as well as to explore the model dependence of the results by using two rather different frameworks for effective nuclear interactions. The Skyrme-type functional SkM* is employed in the nonrelativistic framework, while the density-dependent meson-exchange interaction DD-ME2 is implemented in the relativistic approach. Both the isoscalar and isovector pairing interactions are taken into account within the FT-PNQRPA. Model calculations show that below the critical temperatures the Gamow-Teller excitations display a sensitivity to both the finite temperature and pairing effects, and this demonstrates the necessity for implementing both in the theoretical framework. 
The established FT-PNQRPA opens perspectives for the future complete and consistent description of
astrophysically relevant weak interaction processes in nuclei at finite temperature such as $\beta$-decays, electron capture and neutrino-nucleus reactions.

\end{abstract}


\maketitle

\section{INTRODUCTION} 
Understanding the behavior of nuclei under extreme conditions of isospin and temperature is a long-standing challenge for both theoretical and experimental nuclear physics.
Over the past few decades, the multipole responses in nuclei have been used to probe the properties of nuclei around the valley of stability, and provided a wealth of information about nuclear structure and dynamics. Among various modes of excitation, the spin-isospin response is known as one of the fundamental phenomena in nuclei, studied extensively over the past years \cite{RevModPhys.64.491,ICHIMURA2006446,FUJITA2011549}. These excitation modes are not only important
for nuclear physics, but also for nuclear astrophysics. While the properties of the spin-isospin response can provide valuable information on the spin and spin-isospin dependence of the effective nuclear interaction, detailed knowledge of their properties is relevant in the calculations of the nuclear weak interaction processes in stellar environments (e.g., electron capture, $\beta$-decay, neutrino capture and scattering etc.). It is also known that the nuclear weak interaction processes take place under different conditions of density and temperatures ranging from several hundreds of keV to MeV \cite{RevModPhys.75.819,JANKA200738}. Therefore, the accurate description of the spin-isospin excitations by considering these conditions is of particular relevance to achieve a better
understanding of the behavior of nuclei under extreme conditions and stellar weak interaction processes involving nuclei.

Among the spin-isospin excitations, the Gamow-Teller (GT) transitions have been extensively studied using different theoretical approaches: the shell model \cite{CAURIER1999439,PhysRevC.97.054321,PhysRevC.97.024310} and the relativistic \cite{PhysRevC.69.054303,PhysRevC.95.044301,PhysRevLett.101.122502,PhysRevC.98.051301,LITVINOVA2014307,NIU2013172} and nonrelativistic \cite{PhysRevC.60.014302,PhysRevC.65.054322,PhysRevC.72.064310, PhysRevC.76.044307,PhysRevC.90.054335,PhysRevC.98.024311,Deloncle2017,BAI2013116,PhysRevC.89.044306,NIU2018325} nuclear energy density functionals. 
For more details on previous studies of the GT transitions within these and other theoretical frameworks, see review articles \cite{BROWN2001517,RevModPhys.77.427,Paar_2007} and references therein. 
The applicability of the shell model is limited to medium mass nuclei (A$\leq$ 70), whereas the relativistic and nonrelativistic nuclear energy density functionals have the advantage of describing the excitation properties of nuclei in a consistent approach along the nuclide chart. Presently, one of the open issues in the description of the properties of nuclei is the role of the pairing correlations, which can lead to important modifications in the ground-state properties and respective response functions of nuclei. In open-shell nuclei, the isovector pairing 
between protons and between neutrons ($T=1, S=0$) contributes in the ground-state calculations and it is responsible for the partial occupation of the single-particle states. 
The strength of the isovector pairing is usually adjusted to the empirical pairing gaps, calculated from the nuclear masses using the three points formula \cite{Bender2000,CHANGIZI2015210}. 
In addition to the isovector pairing, the isoscalar {\it proton-neutron} pairing contributes at the level of the residual interaction of the proton-neutron quasiparticle random phase approximation (PNQRPA). 
The significance of the proton-neutron pairing correlations in nuclear structure,
as well as the competition between the isoscalar and isovector pairing modes have already been extensively studied using various approaches (see e.g. \cite{PhysRevC.60.014311, LANGANKE1997253,MARTINEZPINEDO1999379,POVES1998203,warner2006, SAMBATARO2015137, PhysRevC.98.064319,FRAUENDORF201424,Sagawa_2016} and references therein).
The effect of isoscalar proton-neutron pairing has also been discussed in the
case of GT transitions in several energy density functional frameworks \cite{PhysRevC.69.054303,PhysRevC.72.064310,PhysRevC.76.044307,BAI2013116,PhysRevC.90.054335,PhysRevC.95.044301,PhysRevC.60.014302,PhysRevC.98.024311,NIU2018325}. It was shown that the isoscalar pairing reduces the excitation energies, and leads to an enhancement of the low-energy strength (see also Ref. \cite{Sagawa_2016} and the references therein). Furthermore, its inclusion affects the predictions for the GT excitations and $\beta$-decay rates of nuclei \cite{PhysRevC.60.014302,NIU2018325,NIU2013172}. These studies also indicate that the strength of the isoscalar pairing should be slightly larger than or equal to the isovector pairing strength \cite{PhysRevC.90.054335,Sagawa_2016}. However, extensive studies are still needed to constrain the isoscalar pairing strength at zero temperature.

The temperature effects on the ground-state properties \cite{GOODMAN198130,KHAN200794,PhysRevC.88.034308,refId0,PhysRevC.92.014302,PhysRevC.96.024304,PhysRevC.97.054302} and excitations \cite{SOMMERMANN1983163,KHAN2004311,PhysRevC.96.024303,PhysRevC.63.032801,PhysRevC.81.015804,NIU2009315,PhysRevC.63.034323,PhysRevC.83.045807,PhysRevC.80.055801,PhysRevLett.121.082501,PhysRevC.94.015805} in nuclei have also been the subject of several studies, not only to understand their properties under extreme conditions, but also in relation to their relevance for astrophysical processes. Long ago, the effect of temperature on the stellar electron capture rates was studied in neutron-rich germanium isotopes using the hybrid model composed of the shell model Monte Carlo (SMMC) approach and the random
phase approximation (RPA) \cite{PhysRevC.63.032801}. The findings of this study indicate that the configuration mixing and thermal effects can unblock the allowed GT transitions. In Ref. \cite{PhysRevC.81.015804}, the GT$^{+}$ strength distributions and electron capture rates were calculated at finite temperature, based on the PNQRPA and thermo-field-dynamics formalism. It was shown that the GT$^{+}$ excitation energies decrease due to the thermal effects. The relativistic \cite{PhysRevC.83.045807} and nonrelativistic \cite{PhysRevC.80.055801} finite temperature RPA calculations were also performed to study the electron capture on nuclei in stellar environment, using the relevant charge-exchange excitations at finite temperatures. 
Although the calculations were performed for open-shell nuclei, the pairing correlations were not taken into account.
Recently, the nuclear charge-exchange excitations were studied using the finite temperature relativistic nuclear field theory framework; in particular, $\beta$-decay rates have been studied for the first time by going beyond the one-loop approximation \cite{lit18}. However, these calculations were also limited to the closed shell nuclei.
At finite temperatures, the proper description of the spin-isospin excitations in open-shell nuclei requires the inclusion of pairing correlations both in the isovector and isoscalar channels, assuming that the temperature is below the
critical value for the pairing collapse. Therefore, extensions of the current theoretical models are necessary to study the excitations in open-shell nuclei at finite temperatures. For this purpose, the finite temperature QRPA was developed and self-consistent calculations were performed using nonrelativistic functionals to study the electric dipole and quadrupole excitations in nuclei with increasing temperature \cite{PhysRevC.96.024303}. In this work, the finite temperature proton-neutron QRPA (FT-PNQRPA) based on the relativistic and nonrelativistic functionals is developed to study the spin-isospin excitations in open-shell nuclei at finite temperatures. Using the FT-PNQRPA, one can explore the behavior of nuclei under extreme conditions as well as the interplay between the pairing and temperature effects, which is crucial for the proper description of the spin-isospin excitations, especially at temperatures below 1 MeV. With this aim, the relativistic and nonrelativistic nuclear energy density functionals are employed to study the GT$^{-}$ response of open-shell $^{42}$Ca, $^{46}$Ti, and $^{118}$Sn nuclei at zero and finite temperatures. We focus on the properties of the GT$^{-}$ excitations under the influence of both the finite temperature and pairing correlation effects. 

The paper is organized as follows. In Sec. \ref{model}, the formalism of the FT-PNQRPA
is introduced. In Sec. \ref{results}, the results of the calculations are presented. The effects of the isoscalar pairing and temperature on the  GT$^{-}$ excitations of the selected nuclei are studied. The calculations are performed mainly below the critical temperatures, for which pairing correlations still play a role. The competition between the pairing and temperature effects is discussed for the Gamow-Teller Resonance (GTR) and low-energy excitations. Finally, the conclusions and outlook are given in Sec. \ref{finito}.

\section{MICROSCOPIC MODEL: THE FINITE TEMPERATURE PROTON-NEUTRON QRPA}
\label{model}
In the present work, model calculations are carried out using the relativistic and nonrelativistic nuclear energy density functionals, assuming spherical symmetry. In the nonrelativistic framework, the ground-state properties of nuclei are described using the finite temperature Hartree-Fock Bardeen-Cooper-Schrieffer (FT-HFBCS) calculation with Skyrme-type functional SkM* \cite{SKM}. In the relativistic framework, the finite temperature Hartree BCS (FT-HBCS) calculations are performed using the density dependent meson-exchange DD-ME2 functional \cite{DD-ME2}. Detailed information about the the finite temperature H(F)BCS theory can be found in Refs. \cite{refId0,GOODMAN198130}. In the finite temperature framework, the occupation probabilities of the states are given by
\begin{equation}
n_i=v_{i}^{2}(1-f_{i})+u_{i}^{2}f_{i},
\end{equation} 
where $u_i$ and $v_i$ are the BCS amplitudes. The temperature dependent Fermi-Dirac distribution function is given by
\begin{equation}
\label{eq:fd}
f_{i}=[1+\exp(E_{i}/k_{B}\text{T})]^{-1},
\end{equation}
where $E_{i}$ is the quasiparticle (q.p.) energy, $k_{B}$ is the Boltzmann constant, and T is the temperature.

In open-shell nuclei, the isovector pairing ($T=1, S=0$) contributes in the ground-state calculations and leads to the partial occupation of states, while the isoscalar pairing ($T=0, S=1$) contributes only to the residual proton-neutron particle-particle interaction at the FT-PNQRPA level. We note that the isoscalar proton-neutron pairing cannot be considered within the present H(F)BCS framework used for the ground-state calculations since its inclusion represents a rather complex problem.
In the nonrelativistic model, the zero-range density-dependent surface pairing interaction is used in both the isovector and isoscalar pairing channels. The isovector pairing interaction is given by
\begin{equation}
V_{iv}({\bf r_1,r_2})=-V_0^{iv}\frac{1-P_{\sigma}}{2}\left(1-\frac{\rho({\bf r})}{\rho_o}\right)\delta({\bf r_1- r_2}),
\label{T1-pair}
\end{equation}
where $\rho_{0}=0.16$ fm$^{-3}$ and P$_{\sigma}$ is the spin exchange operator. The isovector pairing strength ($V_0^{iv}$) is adjusted to the empirical pairing gap values for the considered nuclei. The isoscalar pairing interaction is given by
\begin{equation}
V_{is}({\bf r_1,r_2})=-V_0^{is}\frac{1+P_{\sigma}}{2}\left(1-\frac{\rho({\bf r})}{\rho_o}\right)\delta({\bf r_1-r_2}),
\label{T0-pair1}
\end{equation}
where $V_0^{is}$ denotes the strength of the isoscalar pairing interaction. Since this channel of the pairing interaction cannot be constrained at the level of
the ground-state, it is parameterized separately within the PNQRPA, e.g.,
by constraining its strength to the properties of GT excitations or $\beta$-decay half-lives. For the purpose of the present analysis $V_0^{is}$ is used as a free parameter in order to explore the model dependence of the
GT$^{-}$ excitation properties at finite temperature.

In the relativistic approach, the isovector and isoscalar pairing are treated differently. For the finite temperature Hartree BCS calculations, we use a monopole pairing interaction \cite{GOODMAN198130}, in which the isovector pairing strength ($G_{0}^{iv}$) is adjusted to the empirical pairing gaps. We also introduce the smooth energy-dependent cut-off weights to take into account the finite range of the pairing interaction (see Refs. \cite{GOODMAN198130,Bender2000} for more information). For the isoscalar pairing, we employ formulation with a short range repulsive Gaussian combined with a weaker longer range attractive Gaussian
\begin{equation}
V_{12}=-G_0^{is}\sum_{j=1}^{2}g_{j}e^{-r_{12}^{2}/\mu_{j}^{2}}\prod_{S=1,T=0},
\label{T0-pair2}
\end{equation}
where $\prod_{S=1,T=0}$ projects onto states with $S=1$ and $T=0$. The ranges $\mu_{1}$=1.2 fm and $\mu_{1}$=0.7 fm of the two Gaussians are from the Gogny interaction, and the relative strengths are set as $g_1$=1 and $g_2=-2$ so that the force is repulsive at small distances \cite{PhysRevC.60.014302,PhysRevC.69.054303}. The residual isoscalar pairing strength ($G_0^{is}$) is also taken as a free parameter that can be constrained by the experimental data at the level of PNQRPA calculations. Note that, since the isoscalar pairing force adopted here is different from the isovector one, we cannot directly compare the relative strength between isovector pairing and isoscalar pairing as in the nonrelativistic framework.

As mentioned above, the FT-PNQRPA is applied on top of the FT-H(F)BCS calculation to describe the excited states of nuclei. The finite temperature PNQRPA matrix is given by
\begin{equation}
\left( { \begin{array}{cccc}\label{eq:qrpa}
 \widetilde{C} & \widetilde{a} & \widetilde{b} & \widetilde{D} \\
 \widetilde{a}^{+} & \widetilde{A} & \widetilde{B} & \widetilde{b}^{T} \\
-\widetilde{b}^{+} & -\widetilde{B}^{\ast} & -\widetilde{A}^{\ast}& -\widetilde{a}^{T}\\
-\widetilde{D}^{\ast} & -\widetilde{b}^{\ast} & -\widetilde{a}^{\ast} & -\widetilde{C}^{\ast}
 \end{array} } \right)
 \left( {\begin{array}{cc}
\widetilde{P}  \\
\widetilde{X }  \\
\widetilde{Y}  \\
\widetilde{Q} 
 \end{array} } \right)
 = E_{\nu}
  \left( {\begin{array}{cc}
\widetilde{P}  \\
\widetilde{X}  \\
\widetilde{Y}  \\
\widetilde{Q} 
\end{array} } \right). \end{equation}
Here, $ E_{\nu}$ represents the eigenvalues of the matrix after the diagonalization, and the eigenvectors are denoted by $\widetilde{P}, \widetilde{X}, \widetilde{Y}$, and  $\widetilde{Q}$. The FT-PNQRPA matrices are diagonalized in a self-consistent way, providing a state-by-state analysis for each excitation. The temperature dependencies of the matrices are given by \cite{SOMMERMANN1983163,PhysRevC.96.024303}
\begin{align}
\begin{split}
\widetilde{A}_{abcd}=&\sqrt{1-f_{a}-f_{b}} A'_{abcd}\sqrt{1-f_{c}-f_{d}}\\
&+(E_{a}+E_{b})\delta_{ac}\delta_{bd} \label{eq:tempa},
\end{split}\\
\begin{split}
\widetilde{B}_{abcd}=&\sqrt{1-f_{a}-f_{b}} B_{abcd}\sqrt{1-f_{c}-f_{d}}, 
\end{split}\\
\begin{split}
\widetilde{C}_{abcd}=&\sqrt{f_{b}-f_{a}} C'_{abcd}\sqrt{f_{d}-f_{c}}\\
&+(E_{a}-E_{b})\delta_{ac}\delta_{bd}\label{eq:tempc},
\end{split}\\
\begin{split}
\widetilde{D}_{abcd}=&\sqrt{f_{b}-f_{a}} D_{abcd}\sqrt{f_{d}-f_{c}}, 
\end{split}\\
\begin{split}
\widetilde{a}_{abcd}=&\sqrt{f_{b}-f_{a}} a_{abcd}\sqrt{1-f_{c}-f_{d}}, 
\end{split}\\
\begin{split}
\widetilde{b}_{abcd}=&\sqrt{f_{b}-f_{a}} b_{abcd}\sqrt{1-f_{c}-f_{d}}, 
\end{split}\\
\begin{split}
\widetilde{a}_{abcd}^{+}=&\widetilde{a}_{abcd}^{T}=\sqrt{f_{d}-f_{c}} a_{abcd}^{+}\sqrt{1-f_{a}-f_{b}}, 
\end{split}\\
\begin{split}
\widetilde{b}_{abcd}^{T}=&\widetilde{b}_{abcd}^{+}=\sqrt{f_{d}-f_{c}} b_{abcd}^{T}\sqrt{1-f_{a}-f_{b}},\label{eq:temp1}
\end{split}
\end{align}
where $E_{a(b)}$ is the quasiparticle energy of either proton($p$) or neutron($n$) states obtained from the ground-state calculations. It should be noted that the diagonal part of the FT-PNQRPA matrix includes both
($E_{a}+E_{b}$) and ($E_{a}-E_{b}$) configuration energies. To guide the reader in the rest of the paper, we only provide the explicit forms of the diagonal matrix elements of the FT-PNQRPA. The $A'$ and $C'$ read
\begin{gather}
\begin{aligned} \label{A}
A'_{abcd}&=(u_{a}u_{b}u_{c}u_{d}+v_{a}v_{b}v_{c}v_{d})V^{\text{pp}}_{abcd} \\
&+(u_{a}v_{b}u_{c}v_{d}+v_{a}u_{b}v_{c}u_{d})V^{\text{ph}}_{a\bar{d}\bar{b}c} \\
&-(-1)^{j_{c}+j_{d}+J}(u_{a}v_{b}v_{c}u_{d}+v_{a}u_{b}u_{c}v_{d})V^{\text{ph}}_{a\bar{c}\bar{b}d},
\end{aligned} \\
\begin{aligned} \label{C}
C'_{abcd}&=(u_{a}v_{b}u_{c}v_{d}+v_{a}u_{b}v_{c}u_{d})V^{\text{pp}}_{a\bar{b}c\bar{d}} \\
&+(u_{a}u_{b}u_{c}u_{d}+v_{a}v_{b}v_{c}v_{d})V^{\text{ph}}_{adbc} \\
&+(-1)^{j_{c}+j_{d}+J}(u_{a}u_{b}v_{c}v_{d}+v_{a}v_{b}u_{c}u_{d})V^{\text{ph}}_{a\bar{c}b\bar{d}},  
\end{aligned} 
\end{gather}
where $V^{\text{ph}}$ and $V^{\text{pp}}$ represent the residual proton-neutron particle-hole ($ph$) and particle-particle ($pp$) interactions, respectively, and the bar denotes time-reversal. In the FT-PNQRPA matrix the $\widetilde{A}$ and
$\widetilde{B}$ matrices describe the effects of the excitations of
quasiparticle pairs, which also contribute at zero temperature.
The other components of the FT-PNQRPA matrix, $\widetilde{C}$, $\widetilde{D}$,
$\widetilde{a}$, $\widetilde{b}$, $\widetilde{a}^{+}$, and
$\widetilde{b}^{T}$ start to play a role at finite temperature because they depend on the modifications of the occupation factors. Detailed information about the other matrices can be found in Ref. \cite{PhysRevC.96.024303}. The FT-PNQRPA amplitudes read
\begin{gather}
\begin{aligned}
\widetilde{X}_{ab}=X_{ab}\sqrt{1-f_{a}-f_{b}},
\end{aligned} \\
\begin{aligned}
\widetilde{Y}_{ab}=Y_{ab}\sqrt{1-f_{a}-f_{b}},
\end{aligned} \\
\begin{aligned}
\widetilde{P}_{ab}=P_{ab}\sqrt{f_{b}-f_{a}},
\end{aligned}\\
\begin{aligned}
\widetilde{Q}_{ab}=Q_{ab}\sqrt{f_{b}-f_{a}}.
\end{aligned}
\end{gather}

In the present work, the structure of the excited states is also analyzed using the FT-PNQRPA amplitudes.
For a given excited state $ E_{\nu}$, the contribution of the quasiparticle configurations to the excitation is determined by 
\begin{equation}
A_{ab}=|\widetilde{X}_{ab}^{\nu}|^{2}-|\widetilde{Y}_{ab}^{\nu}|^{2}+|\widetilde{P}_{ab}^{\nu}|^{2}-|\widetilde{Q}_{ab}^{\nu}|^{2},
\label{Aab}
\end{equation} 
and the normalization condition can be written as
\begin{equation}
\sum_{a>b}A_{ab}=1.
\label{aaa}
\end{equation}
At finite temperatures, the GT$^{-}$ strength for the $n\rightarrow p$ transitions ($E_{p}>E_{n}$) is calculated using
\begin{equation}
\begin{split}
B(\text{GT}^{-})&=\bigl|\langle \nu ||\hat{F}_{J}||\widetilde0\rangle \bigr|^{2}\\
&=\biggl|\sum_{p>n}\Big\{(\widetilde{X}_{pn}^{\nu}u_{p}v_{n}+ \widetilde{Y}_{pn}^{\nu}v_{p}u_{n})\sqrt{1-f_{n}-f_{p}} \\
&+(\widetilde{P}_{pn}^{\nu}u_{p}u_{n}-\widetilde{Q}_{pn}^{\nu}v_{p}v_{n})\sqrt{f_{n}-f_{p}}\Big\}\langle p ||\hat{F}_{J}||n\rangle\biggr|^{2},
\end{split}
\label{bel}
\end{equation}
where $|\nu\rangle$ is the excited state and $|\widetilde0\rangle$ is the correlated FT-PNQRPA ground-state. The $\text{GT}^{-}$ transition operator reads $\hat{F}_{J}={\bm{\sigma}}\tau_{-}$. In the FT-PNQRPA calculations, the quasiparticle energy cutoff is taken as E$_{cut}$ =100 MeV in order to ensure the convergence of the results.

\section{RESULTS}
\label{results}
In this section, the FT-PNQRPA is employed to
study the pairing and temperature effects in the
GT$^{-}$ transition strengths for $^{42}$Ca, $^{46}$Ti, and $^{118}$Sn. Both the relativistic and nonrelativistic nuclear energy density functionals are employed in the calculations. In the nonrelativistic framework, the Skyrme-type SkM* interaction \cite{SKM} is used in the calculations due to its success in the description of the Gamow-Teller excitations \cite{PhysRevC.94.064328} as well as the $\beta$-decay half lives of nuclei \cite{NIU2018325}. In the relativistic calculations, the meson-exchange density-dependent relativistic mean field effective interaction DD-ME2 is adopted \cite{DD-ME2}. We verify that the Ikeda sum rule \cite{IKEDA1963271} is satisfied at both zero and finite temperatures. In the following, $\pi$ and $\nu$ denote the protons and neutrons, respectively.

\subsection{\texorpdfstring{$^{42}$Ca} {} nucleus}
\label{1}
In this part, we study the effects of the isoscalar pairing and temperature on the GT$^{-}$  excitations in the $^{42}$Ca nucleus. Recently, the low and high-energy collective GT$^{-}$ excitations were observed for $f$-shell nuclei in the high-resolution ($^{3}$He, $t$) measurements, and in the case of $^{42}$Ca it was shown that most of the strength was collected in the low-energy region at 0.61 MeV \cite{PhysRevLett.112.112502,PhysRevC.91.064316}. Then, the effect of the isoscalar pairing on the GT$^{-}$ excitations was studied in nuclei with mass number A=42-58 in order to explain the experimental results \cite{PhysRevC.90.054335}. The results indicate that the inclusion of the isoscalar pairing reduces the GT$^{-}$ excitation energies and leads to an increase of the low-energy strength, thereby the theoretical results become more compatible with the experimental data. Since the GT$^{-}$ excitations are quite sensitive to the isoscalar pairing, it would be interesting to study the competition between the temperature and pairing effects in the $^{42}$Ca nucleus.

\begin{figure*}[!ht]
  \begin{center}
\includegraphics[width=1\linewidth,clip=true]{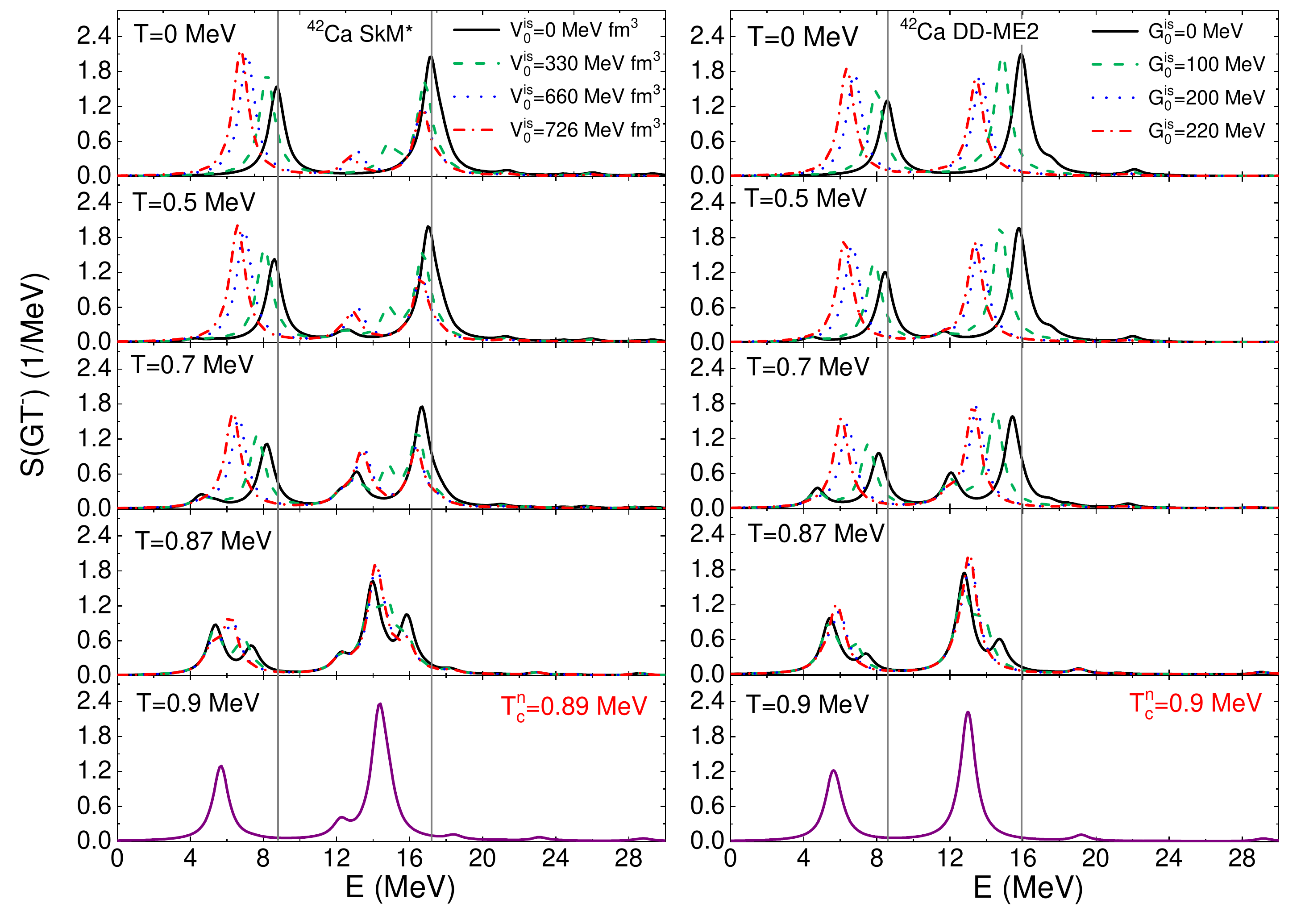}
  \end{center}
 \caption{Left panel: the GT$^{-}$ strength for $^{42}$Ca calculated using the SkM* interaction at T=0, 0.5, 0.7, 0.87 and 0.9 MeV. The isoscalar pairing strength is varied as $V_{0}^{is}$=0, 330, 660 and 726 MeV fm$^{3}$ (see text for explanation). Right panel: the same
but for the DD-ME2 interaction in the particle-hole channel and the finite range isoscalar pairing. The isoscalar pairing strength is varied as $G_0^{is}$=0, 100, 200 and 220 MeV. The excitation energies are calculated with respect to the ground-state of the parent nucleus. The excited states are smoothed with a Lorentzian of width $\Gamma=1$ MeV. The vertical gray lines are drawn to guide the eye.} 
  \label{42}
\end{figure*}

In Fig. \ref{42}, the GT$^{-}$ strength is displayed for $^{42}$Ca using the SkM* (left panels) and DD-ME2 (right panels) functionals.
The excited states are smoothed with a Lorentzian having width $\Gamma=1$ MeV. We note it is an arbitrary value used only for the presentation purposes. Further developments toward including couplings with complex configurations are needed to microscopically calculate the spreading widths to be used in excitation transition strength distributions.
The isovector pairing strength is adjusted to the empirical neutron pairing gap value. In this work, the isovector pairing strengths are taken as $V_{0}^{iv}$=660 MeV fm$^{3}$ and $G_{0}^{iv}$=26 MeV/A for the SkM* and DD-ME2 functionals, respectively. There are no pairing effects for protons due to the shell closure at Z=20. In order to explore the model dependence of the results
on the strength of the isoscalar pairing, calculations are performed by assuming different values for the strength at T=0, 0.5, 0.7, and 0.87 MeV. The isoscalar pairing strength is varied by using different $V_{0}^{is}$ ($G_{0}^{is}$) values in the calculations using the SkM* (DD-ME2) functional [see Eqs. (\ref{T0-pair1}) and (\ref{T0-pair2}) and the relevant discussion in Sec. \ref{model}]. 

\begin{table*}[ht]
\caption{The diagonal matrix elements of the FT-PNQRPA [see Eq. (\ref{eq:qrpa})] and their components for the ($\pi 1f_{7/2}$,$\nu 1f_{7/2}$) configuration in $^{42}$Ca. The calculations are performed using the SkM* functional with increasing temperature. The isoscalar pairing is not taken into account. Herein, $\pi$ and $\nu$ refer to the proton and neutron states, respectively.} \label{table1} 
\begin{tabular}{c c c c c c c} 
\hline\hline  \\[-0.5em]
$^{42}$Ca-SkM*  & T=0.0 MeV& T=0.5 MeV   &T=0.7 MeV&T=0.87 MeV&T=0.9 MeV\\ [1ex]
\hline\\ [-1.ex]
$E_{\pi}+E_{\nu}$ (MeV)     &4.10  & 4.19  & 4.00  & 3.58   & 3.42 \\ 
($uvuv+vuvu$)                                 &0.26  & 0.25  & 0.21  & 0.07   & 0.0 \\ 
$V_{ph}$ (MeV)                                &3.07  & 3.07  & 3.08  & 3.08   & 3.08 \\ 
(1$-f_{\nu}-f_{\pi}$)               &1.0   & 0.97  & 0.87  & 0.73   & 0.68 \\ 
$\widetilde{A}$ matrix (MeV)                                &4.91  & 4.93  & 4.56  & 3.75   & 3.42 \\ 
\hline\\ [-1.ex]     
$E_{\pi}-E_{\nu}$ (MeV)   &0.2  & 0.46  & 0.80  & 1.36   & 1.55 \\ 
($uuuu+vvvv$)                                 &0.74 & 0.75  & 0.80  & 0.92   & 1.0 \\ 
$V_{ph}$ (MeV)                                &3.07  & 3.07  & 3.08  & 3.08   & 3.08 \\ 
($f_{\nu}-f_{\pi}$)                 &0.0  & 0.01  & 0.06  & 0.16   & 0.20 \\ 
$\widetilde{C}$ matrix (MeV)                                &0.2  & 0.50  & 0.95  & 1.83   & 2.17 \\ 
\hline\hline \\ [-1.ex]
\end{tabular}
\end{table*}

We start our analysis of the GT$^{-}$ excitations by varying the isoscalar pairing strength at zero temperature (the topmost panels in Fig. \ref{42}). Without the isoscalar pairing ($V_{0}^{is}$=0 MeV fm$^{3}$ for SkM* and $G_0^{is}$=0 MeV for DD-ME2), the GTRs and low-energy peaks are obtained at 17.13 (15.85) and 8.73 (8.57) MeV using the SkM* (DD-ME2) functional. The GT$^{-}$ peaks are found at different excitation energies due to the effective interactions employed, resulting in differences in the nuclear shell structure and the FT-PNQRPA residual interactions. The effect of the isoscalar pairing is the same for relativistic and nonrelativistic functionals.
By increasing the isoscalar pairing strength, excitation energies and transition strengths decrease in the GTR region, while the excited states start to shift downward and strength increases in the low-energy region. Since the isoscalar pairing is attractive, the excited state energies become smaller by increasing the isoscalar pairing strength. For instance, the GTR and low-energy peaks are obtained at 16.68 (13.44) and 6.75 (6.36) MeV using the SkM* (DD-ME2) functional for the largest value of the pairing strength. We can also analyze the components of the excited states in order to understand the underlying mechanism of increasing the transition strength in the low-energy region.
Without the isoscalar pairing, the low-energy GT$^{-}$ peak is formed with the ($\pi 1f_{7/2},\nu 1f_{7/2}$) two q.p. configuration, for both the SkM* and DD-ME2 functionals. By increasing the isoscalar pairing strength, the low-energy peak still takes an important contribution from the ($\pi 1f_{7/2},\nu 1f_{7/2}$) configuration. In addition, the ($\pi 1f_{7/2},\nu 1f_{5/2}$) and ($\pi 1f_{5/2},\nu 1f_{7/2}$) start to contribute to the low-energy peak and the strength increases due to the coherent contribution of these configurations.
Although the contributions of the ($\pi 1f_{7/2},\nu 1f_{5/2}$) and ($\pi 1f_{5/2},\nu 1f_{7/2}$) configurations are quite low compared to the ($\pi 1f_{7/2},\nu 1f_{7/2}$), these transitions are impacted by the isoscalar pairing due to the $u$ and $v$ factors participating in the corresponding matrix elements [see Eq. \ref{A}], and play an important role in the increase of the low-energy strength. Similar results are also obtained in Ref. \cite{BAI2013116}. As mentioned above, the experimental data indicates that most of the strength is collected below E$<$12 MeV and around 0.61 MeV for the $^{42}$Ca $\rightarrow$ $^{42}$Sc transition. In experiment, the reduced GT$^{-}$ transition strength is B(GT$^{-}$)=2.173(47) for the excited state at 0.61 MeV, while the total strength below E$<$12 MeV is found to be B(GT$^{-}$)=2.7(4) \cite{PhysRevC.91.064316}.
It is known that the QRPA calculations cannot predict observed fragmentation in the strength. In addition, the total experimental strength is overestimated in the calculations using the QRPA. Nonetheless, we can compare our results with the experimental data in a qualitative manner.
To compare the results of QRPA calculations with the experimental data, the excited state energies with respect to the daughter nuclei are obtained by subtracting the experimental binding energy difference of the parent and daughter nuclei from the excited state energies (see Appendix \ref{appendix}). In the calculations, a strong peak is obtained at 1.52 (1.36) MeV with respect to the daughter nucleus, and its B(GT$^{-}$) value is 2.41 (2.02) when using the SkM* (DD-ME2) functional, in the case without isoscalar pairing. The isoscalar pairing strength can be adjusted to obtain the experimentally observed peak energy. We can set it at 0.6 MeV using the SkM* (DD-ME2) functional if $V_{0}^{is}$=482 MeV fm$^{3}$ ($G_0^{is}$=120 MeV). Then, the strength is also increased and without further fine tuning we find a value for the total B(GT$^{-}$) below 12 MeV equal to 2.96 (2.36) for the SkM* (DD-ME2) functional. Thus, the final result is in better agreement with the experimental findings. The role played by isoscalar pairing to bring theory in better harmony with the measurement has been already highlighted in Refs. \cite{PhysRevC.90.054335,Sagawa_2016,PhysRevC.91.064316}. Note that no quenching factor is employed.

Before discussing the temperature effect on the GT$^{-}$ response, we should clarify the modifications of the ground-state properties of nuclei induced by temperature. As mentioned before, a sharp pairing phase transition is expected at critical temperatures due to the grand-canonical description of nuclei. Accordingly, the isovector pairing, which is leading to the partial occupation of the quasiparticle states, vanishes and does not contribute to the FT-PNQRPA matrices above the critical temperatures. For $^{42}$Ca, the critical temperature ($T_{c}$) values for neutrons are obtained at 0.89 and 0.9 MeV for the SkM* and DD-ME2 functionals, respectively. Therefore, the isovector pairing is still active in the calculations below T$<$0.9 MeV. At zero temperature, $^{42}$Ca has partial occupation probabilities for neutron states due to the isovector pairing, while it has a proton shell closure at Z=20. By increasing temperature, the occupation probabilities of the states below the Fermi level start to decrease, while the states above the Fermi level become populated. For instance, at zero temperature the proton $2s_{1/2}$ and $1d_{3/2}$ states are fully occupied, while by increasing the temperature, the occupation probabilities of the proton $2s_{1/2}$ and $1d_{3/2}$ states decrease and the $1f_{7/2}$ state above the Fermi level starts to be populated. Thus, new excitation channels become possible at finite temperature due to the smearing of the Fermi surface. In our work, we performed  calculations at low temperatures and up to the critical temperature T=0.9 MeV, with the aim to study the competition between the temperature and isoscalar pairing on the GT$^{-}$ states. Since the temperature is not high enough to occupy the states in the continuum, it mainly leads to small changes in the single(quasi)-particle energies and occupation probabilities of the states around the Fermi level. Nonetheless, the GT$^{-}$ states are sensitive to the temperature effects as we discuss below.

Next, we consider the effect of the temperature on the GT$^{-}$ excitations in $^{42}$Ca. We start to analyze the excited states without the isoscalar pairing at finite temperatures (see the solid black lines in each panel of Fig. \ref{42}). In this case, only the particle-hole interaction contributes to the FT-PNQRPA matrices, and the residual proton-neutron particle-particle interaction is ignored. As can be seen from Fig. \ref{42}, the effect of the temperature is quite similar in the results for the SkM* and DD-ME2 functionals. Without the isoscalar pairing, both the GTR and low-energy peaks are shifted downwards and new excited states are obtained with increasing temperature. 
These changes in the GT$^{-}$ states are more apparent for the calculations close to the critical temperatures (i.e., 0.7$\leq$T$<$T$_c$ MeV) due to the rapid decrease of the isovector pairing correlations. By increasing temperature, the strengths of the main GT$^{-}$ peaks decrease and the excited states shift downward due to (1) the decrease in the two q.p. energies and change in the $u$ and $v$ factors of the states as well as (2) the weakening of the repulsive residual particle-hole interaction because of the temperature factors in the matrices. We can better understand the effect of temperature on the GT$^{-}$ excitations by following the changes in the diagonal matrix elements [see Eqs. (\ref{eq:tempa}), (\ref{eq:tempc}), (\ref{A}), and (\ref{C})] of the corresponding two q.p. configurations.  At zero temperature, the most prominent low-energy peak is obtained at 8.73 MeV, which is composed of the ($\pi 1f_{7/2},\nu 1f_{7/2}$) configuration using the SkM* functional. In Table \ref{table1}, we show the changes in the diagonal matrix element ($\widetilde{A}$ matrix) and its components for this configuration with increasing temperature. As can be seen from Table \ref{table1}, the value of the diagonal matrix element becomes smaller due to the decrease in the unperturbed energy, temperature factor and ($uvuv+vuvu$) value with increasing temperature. Therefore, this low-energy state starts to shift downward and its strength is lowered with increasing temperature. Above the critical temperatures, at T=0.9 MeV, the residual interaction part of the diagonal matrix elements does not contribute to the FT-PNQRPA matrices due to the disappearance of isovector pairing correlations, and this configuration disappears from the $ph$ sector of the $\widetilde{A}$ matrix.

By increasing temperature, new excited states are obtained around 5 MeV for both the SkM* and DD-ME2 functionals. While their strengths start to increase, the excited state energies also start to slightly shift upwards with increasing temperature. For instance, the newly formed low-energy peaks around 5 MeV start to be apparent for T$\geq$ 0.5 MeV, and they are mainly formed with the ($\pi 1f_{7/2},\nu 1f_{7/2}$) configuration using both the SkM* and DD-ME2 functionals. These new excited states are formed due to the contribution of the ($E_{\pi}-E_{\nu}$) two q.p. configurations at finite temperatures. Since the energy of this new configuration is small compared to the ($E_{\pi}+E_{\nu}$) energy, the excited states are obtained at lower energies. In Table \ref{table1}, the changes in the corresponding diagonal matrix element ($\widetilde{C}$ matrix) and its components are also given for the SkM* functional for different temperatures. At zero temperature, this configuration does not contribute to the diagonal part of the FT-PNQRPA matrix because of the zero value of the temperature factors. By increasing temperature, the temperature factor of the matrix ($f_{\nu}-f_{\pi}$) no longer zero and it starts to contribute to the matrices. Therefore, the diagonal matrix elements start to play a role in the FT-PNQRPA and the formation of a new low-energy state is obtained at around 5 MeV. This newly formed excited state continues to shift slightly upwards and its strength becomes larger due to the increase of the unperturbed energy as well as the increase in the temperature factor and ($uuuu+vvvv$) value. At T=0.9 MeV, we obtain a single peak in the low-energy region, which is mainly composed of the ($\pi 1f_{7/2},\nu 1f_{7/2}$) configuration and ($E_{\pi}-E_{\nu}$) two q.p. energy in the FT-PNQRPA matrix.

The same physical mechanism is also present in the GTR region with increasing temperature. In this case, the GTR is mainly composed of the ($\pi 1f_{5/2},\nu 1f_{7/2}$) configuration. While the strength and excitation energy of the GTR start to decrease, the strength and excitation energy of the newly formed excited states around 13 MeV increase with increasing temperature. For 0.7$\leq$T$<$0.9 MeV, several peaks are obtained with comparable strengths both in the GTR and low-energy region. Above the critical temperatures, these peaks combine and two strong peaks are obtained, e.g., at 5.67 and 14.33 MeV for the SkM* functional. Similar results are also obtained using the DD-ME2 functional.

\begin{figure}[!ht]
  \begin{center}
\includegraphics[width=1\linewidth,clip=true]{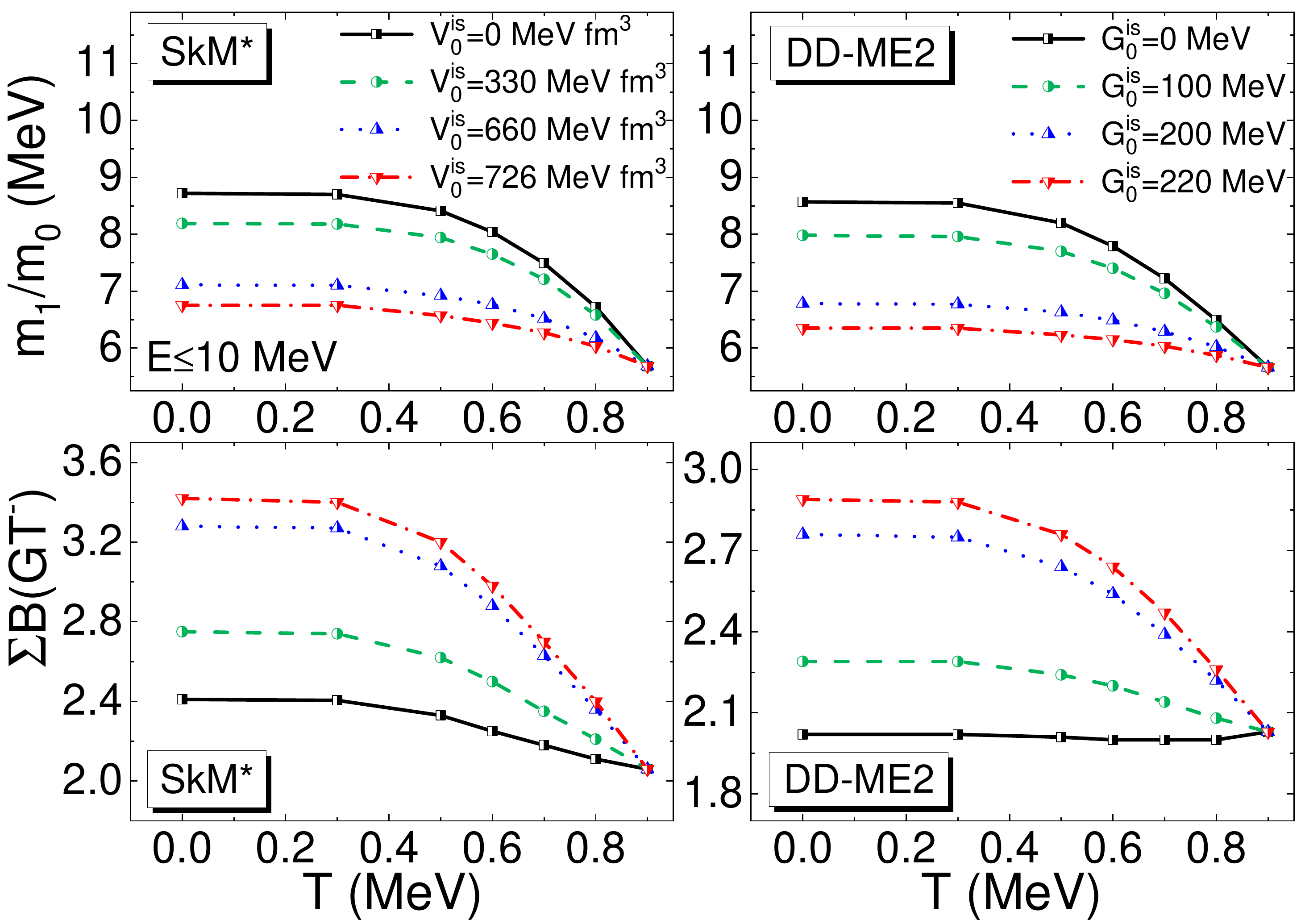}
  \end{center}
 \caption{The centroid energies ($m_1/m_0$) (upper panels) and the total sum of the GT$^{-}$ strengths (lower panels) for the excited states below 10 MeV as a function of temperature for $^{42}$Ca. The calculations are performed using the SkM* (left panels) and DD-ME2 functionals (right panels) and by varying the isoscalar pairing strength.} 
  \label{422}
\end{figure}

In the following, we discuss the behavior of the GT$^{-}$ excitations under the influence of the temperature and isoscalar pairing. For this purpose, the change in the GT$^{-}$ excitations using various isoscalar pairing strengths is also illustrated in Fig. \ref{42} for different temperatures between 0 and 0.9 MeV. The GT$^{-}$ states are influenced in the same way: the excited state energies decrease and new excited states are obtained under the influence of both the isoscalar pairing and temperature effects. However, we conclude that the predictions for the strength and excitation energies of the GT$^{-}$ states depend on the strength of the isoscalar pairing below the critical temperatures, as explained below.

In Fig. \ref{422}, the centroid energies ($m_1/m_0$) \cite{PhysRevC.96.024303} and the total sums of the GT$^{-}$ strengths are displayed for the excited states below 10 MeV as a function of temperature. The calculations are performed with the SkM* (left panels) and DD-ME2 (right panels) functionals using different isoscalar pairing strength values in order to gain a better insight into the effects of the interplay between the temperature and isoscalar pairing in the low-energy region. As mentioned before, inclusion of the isoscalar pairing decreases the excitation energy in the low-energy region due to its attractive nature. In addition, the low-energy strength increases with the contribution of the ($l=l',j=j'\pm$1) as well as the ($l=l',j=j'$) two q.p. configurations in a coherent way. Therefore, the lowest centroid energy and the highest strength values are obtained using the largest isoscalar pairing strength at zero temperature.
By increasing temperature, it is seen that the predictions for the centroid energies and the total sum of the strength is also quite different with and without isoscalar pairing. Without the isoscalar pairing, the centroid energies decrease rapidly after T$>$0.5 MeV, whereas the total strength only slightly changes with increasing temperature. With the inclusion of the isoscalar pairing, an opposite trend is obtained: the temperature leads to a sharp decrease in the total sum of the strength close to the critical temperatures, whereas the centroid energy is slightly lowered. The smooth decrease in the centroid energy, which also takes place in the GTR region, is related to the competition between the temperature and the isoscalar pairing effects.
Without the isoscalar pairing, the centroid energy and strength gradually decreases due to the softening of the repulsive residual $ph$ interaction as well as the decreasing effect of the isovector pairing with increasing temperature. 
In the presence of the isoscalar pairing, the temperature affects both the $ph$ and $pp$ residual interaction matrix elements, and decrease their impact. The effect of isoscalar pairing is to lower the GT$^{-}$ excitation energy and increase the low-lying GT$^{-}$ strength. The effect of isovector pairing is to increase the GT$^{-}$ excitation energy, with little influence on the GT$^{-}$ strength. When temperature is increased, both isovector pairing and isoscalar pairing effects are weakened. As a result, the temperature effect on the  GT$^{-}$ energy is partly canceled due to the opposite effects of isoscalar pairing and isovector pairing, leading to a slow change of the GT$^{-}$ energy as a function of temperature. The GT$^{-}$ strength is instead reduced markedly due to the weakening of isoscalar pairing. Then, we analyze in detail how the isoscalar pairing affects the low-lying  GT$^{-}$ strength based on the numerically calculated wave function (configurations) of the low-lying GT$^{-}$ states.
Albeit small in percentage compared to the ($l=l',j=j'$) configurations, the contribution of the ($l=l',j=j'\pm$1) configurations to the low-energy states in the presence of the isoscalar pairing increases the low-energy strength significantly, as mentioned above. In other words, the low-energy strength is sensitive to the changes related to the ($l=l',j=j'\pm$1) configurations. We find that the contribution of the ($\pi 1f_{5/2},\nu 1f_{7/2}$) and ($\pi 1f_{7/2},\nu 1f_{5/2}$) configurations to the low-energy states start to decrease gradually for T$>$0.7 MeV, due to the weakening of isoscalar pairing. Therefore, the low-energy strength is reduced sharply by increasing temperature. Above the critical temperatures, the pairing properties are washed out. At T=0.9 MeV, ($l=l',j=j'\pm$1) transitions do not contribute to the low-energy states and the excited state at 5.67 MeV is composed of ($\pi 1f_{7/2},\nu 1f_{7/2}$) (87.82\%), ($\pi 2s_{1/2},\nu 2s_{1/2}$) (8.33\%), and ($\pi 1d_{3/2},\nu 1d_{3/2}$) (2.93\%) configurations for the SkM* functional. Similar results are also obtained using the DD-ME2 functional. Considering the results of the present investigation of $^{42}$Ca, it is seen that the inclusion of the pairing correlations in open-shell nuclei plays an important role in the description of the Gamow-Teller excitations below the critical temperatures. 

Before ending this part, we should mention that the FT-PNQRPA calculations are performed by assuming a grand-canonical description of nuclei, which leads to the sharp pairing phase transitions in nuclei at critical temperatures. However, the nucleus is a finite system and the statistical fluctuations should also be taken into account
in the calculations. Recently, this issue was studied using
different theoretical approaches. It was shown that the
pairing phase transitions become smoother and pairing
correlations continue even above T$>$1 MeV \cite{PhysRevC.29.1887, PhysRevLett.61.1926, PhysRevC.76.064320, PhysRevC.47.606, PhysRevC.88.034324}.
Considering the sensitivity of the GT$^{-}$ excitations to the pairing effects in $^{42}$Ca and $^{46}$Ti nuclei, this correction can be important for the calculations of the GT$^{-}$ excitations, especially below T$<$1 MeV. Nonetheless, our model can also be considered as a first step for this kind of calculations. 

\subsection{\texorpdfstring{$^{46}$Ti} {} nucleus}
\label{3}
\begin{figure*}[!htb]
  \begin{center}
\includegraphics[width=1\linewidth,clip=true]{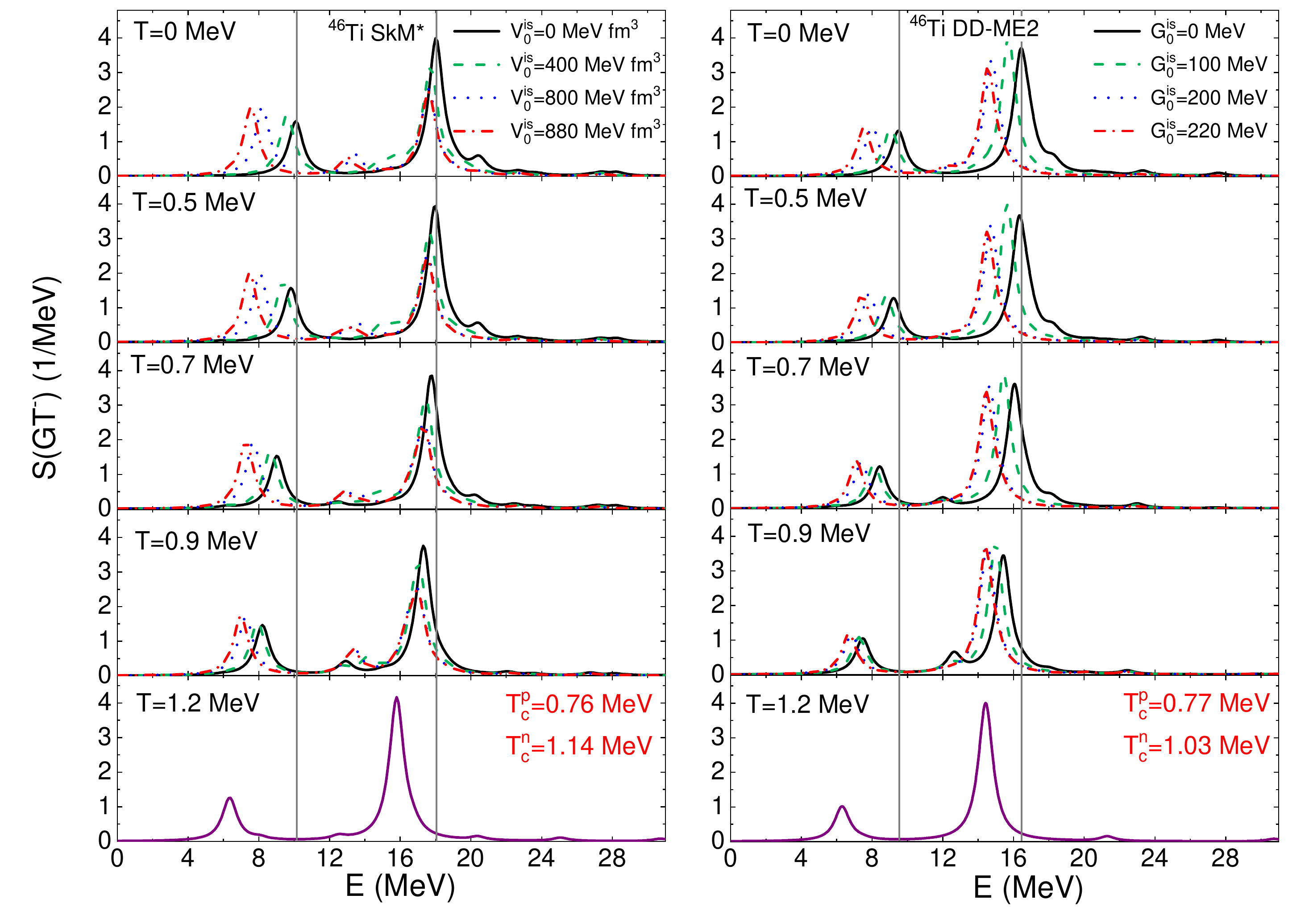}
  \end{center}
 \caption{The same as in Fig. \ref{42}, but for $^{46}$Ti.} 
  \label{46}
\end{figure*}
\begin{table*}
\caption{The centroid energies ($m_1/m_0$) of the GTRs for $^{46}$Ti with and without isoscalar pairing at finite temperatures. The calculations are performed using the SkM* and DD-ME2 functionals between 12.5 and 20.5 MeV.} \label{table2}
\begin{center}
\begin{tabular}{@{\extracolsep{4pt}}ccccccccc@{}}
\hline\hline \\ [-1.5ex]
 && \multicolumn{2}{c}{SkM*} & &\multicolumn{2}{c}{DD-ME2}  \\ \cline {3-4} \cline{6-7} \\[-1.ex]
 & GTR   &$V_{0}^{is}=0$ MeV fm$^{3}$ & $V_{0}^{is}=880$ MeV fm$^{3}$&& $G_{0}^{is}=0$ MeV & $G_{0}^{is}=220$ MeV  \\[0.5em]
\hline   \\[-0.5em]
 & T=0 MeV   &   18.21  & 16.89   &&  16.69  & 14.63			  \\
 & T=0.5 MeV &   18.13  & 16.82   &&  16.57  & 14.61				\\
 & T=0.7 MeV &   17.86  & 16.59   &&  16.24  & 14.58				\\
 & T=0.9 MeV &   16.95  & 16.18   &&  15.18  & 14.54					\\
 & T=1.1 MeV &   16.06  & 15.87   &&  14.43  & 14.43					\\
 & T=1.2 MeV &   15.86  & 15.86   &&  14.43  & 14.43					\\
\hline\hline
\end{tabular}
\end{center}
\vspace{-7mm}
\end{table*}

Recently, the effect of the isoscalar pairing on the GT$^{-}$ excitations in $^{46}$Ti was studied using the PNQRPA with nonrelativistic functionals \cite{PhysRevC.90.054335}. Similar to the findings for $^{42}$Ca, it was shown that the excited state energies shift downward due to the attractive nature of the isoscalar pairing, while the low-energy (GTR) strength increases (decreases) \cite{PhysRevC.90.054335}. The experimental data also indicate that most of the strength is collected below 4 MeV with a strong peak at 0.994 MeV with respect to the daughter nucleus \cite{PhysRevLett.112.112502}. In this part, we discuss the behavior of the GT$^{-}$ excitations in $^{46}$Ti under the influence of temperature and isoscalar pairing. In this way, we can test our understanding using a nucleus in which both neutrons and protons are sensitive to pairing. Note that we include 
isovector pairing for both species, so that both proton and neutron states are partially occupied in $^{46}$Ti at zero temperature. 

For $^{46}$Ti, the isovector pairing strength is taken as $V_{0}^{iv}$=800 MeV fm$^{3}$ for neutrons and protons using the SkM* functional. For the calculations using the DD-ME2 functional, the isovector pairing strength $G_{0}^{iv}$ is taken as 28 and 26 MeV/A for neutrons and protons, respectively. The critical temperature values for neutrons(protons) are obtained at T$^{n(p)}_{c}$=1.14(0.76) and 1.03(0.77) MeV using the SkM* and DD-ME2 functionals, respectively. Since the proton and neutron states are already partially occupied at zero temperature, the temperature mainly changes the single(quasi)-particle energies and occupation probabilities of the $sd$ and $pf$ shells around the Fermi level.

In Fig. \ref{46}, the GT$^{-}$ excitations are displayed for various isoscalar pairing strengths at finite temperatures. We start our analysis by increasing the isoscalar pairing strength at zero temperature (the top most panels). Similar to the findings in Sec. \ref{1}, the low-energy states start to shift downwards and the strength increases slightly, while the GTR energy and strength decreases with increasing isoscalar pairing strength. Without the isoscalar pairing at zero temperature, the GTR peaks are obtained at 18.0 and 16.41 MeV using the SkM* and DD-ME2 functionals, respectively. 
In addition, the GTR is mainly composed of the ($\pi 1f_{5/2},\nu 1f_{7/2}$) configuration. With the inclusion of the isoscalar pairing for the SkM* functional, the ($\pi 1f_{7/2},\nu 1f_{5/2}$) and ($\pi 2p_{3/2},\nu 2p_{1/2}$) configurations start to contribute to the GTR peak. For the DD-ME2 functional, these configurations are also accompanied by the ($\pi 2p_{1/2},\nu 2p_{3/2}$) and ($\pi 2p_{3/2},\nu 2p_{3/2}$). Nonetheless, the main contribution to the GTRs still comes from the ($\pi 1f_{5/2},\nu 1f_{7/2}$) using the SkM* and DD-ME2 functionals. In addition, the GTR energy decreases due to the attractive nature of the isoscalar pairing. For the largest values of the isoscalar pairing strengths, the GTR peaks are obtained at 17.61 and 14.56 MeV using the SkM* and DD-ME2 functionals, respectively. In comparison to the results for the SkM* functional, the decrease in the GTR energy is larger using the DD-ME2 functional due to the lower values of the unperturbed energies of the contributing two q.p. configurations. The low-energy region is also impacted by varying the isoscalar pairing strength. Using the SkM* and DD-ME2 functionals without isoscalar pairing, the low-energy peaks are obtained at 10.1 (2.27) and 9.50 (1.67) MeV with respect to the parent (daughter) nuclei, respectively. Using the largest values of the isoscalar pairing strength, these states are obtained at 7.56 (\mbox-0.27) and 7.55 (-0.28) MeV for the SkM* and DD-ME2 functionals, respectively. We also look for the appropriate isoscalar pairing strength values in order to compare our theoretical results with the experimental data. Using the SkM* (DD-ME2) functional with $V_{0}^{is}$=680 MeV fm$^{3}$ ($G_0^{is}$=132 MeV), we find that the low-energy peak is obtained around 1.0 MeV with respect to the daughter nuclei, and is in good agreement with the experimentally observed peak at 0.994 MeV.
With and without the isoscalar pairing, the low-energy peaks are mainly composed of the ($\pi 1f_{7/2},\nu 1f_{7/2}$) configuration. Similar to the findings in Sec \ref{1}, the ($\pi 1f_{7/2},\nu 1f_{5/2}$) configuration starts to contribute to the low-energy states in a coherent way with the inclusion of the isoscalar pairing. Therefore, the low-energy strength becomes more pronounced.

\begin{table*}
\caption{The excitation energies and the strengths of the relevant peaks in the low-energy region of $^{46}$Ti with increasing temperature. The calculations are performed using the SkM* functional, and the isoscalar pairing strength is taken as $V_{0}^{is}=880$ MeV fm$^{3}$. The quasiparticle configurations and their contribution to the norm of the state (in percentage) [see Eqs. (\ref{Aab}) and (\ref{aaa})] are also provided.} \label{tableqq}
\begin{center}
\begin{tabular}{@{\extracolsep{4pt}}ccccccccc@{}}
\hline\hline \\ [-1.5ex]
 &SkM*          &T=0 MeV & T=0.7 MeV & T=0.9 MeV & T=1.1 MeV& T=1.2 MeV\\ [1.ex] \cline{3-3} \cline{4-4}\cline{5-5} \cline{6-6} \cline{7-7} \\ [-1.5ex] 
 && E=7.56 MeV &E=7.27 MeV&E=7.0 MeV &E=6.57 MeV   &E=6.36 MeV\\ 
 &Configurations& B(GT$^{-}$)=3.12 & B(GT$^{-}$)=3.08& B(GT$^{-}$)=2.68  & B(GT$^{-}$)=2.07   & B(GT$^{-}$)=1.94 \\ 
\hline
 & ($\pi 1f_{7/2},\nu 1f_{7/2}$) & 90.18  & 90.41  & 91.46 & 94.75 & 89.90  				\\
 & ($\pi 1f_{7/2},\nu 1f_{5/2}$) & 2.63   & 4.22   & 3.90  & 1.10   &   				\\
 & ($\pi 1f_{5/2},\nu 1f_{7/2}$) & 0.51   & 1.64   & 2.14  &       &   				\\
 & ($\pi 1g_{9/2},\nu 1g_{9/2}$) & 0.96   & 1.22   & 1.14  &        &   				\\
 & ($\pi 2p_{3/2},\nu 2p_{3/2}$) & 1.75   & 1.77   & 1.66  & 1.33 & 2.94  				\\
 & ($\pi 2s_{1/2},\nu 2s_{1/2}$) &        &        &       & & 5.95				\\
\hline\hline
\end{tabular}
\end{center}
\vspace{-7mm}
\end{table*}
As expected, the temperature effects decrease the energies of the GT$^{-}$ states. Similar to the findings in Sec. \ref{1}, this decrease is more pronounced close to the critical temperatures due to the rapid change of the ground-state properties [i.e., $u$ and $v$ factors, and single(quasi)-particle energies of states] as well as the weakening of the residual interaction. In case of no isoscalar pairing, it is also seen that the strength of the new low-energy peak is negligible in $^{46}$Ti as compared with the results of $^{42}$Ca at finite temperatures.
In Table \ref{table2}, we compare the centroid energies of the GTRs with and without isoscalar pairing at finite temperatures. The centroid energies are calculated between 12.5 and 20.5 MeV to display the changes in the GTR region. Compared to the results without isoscalar pairing, the centroids of the GTRs are obtained at lower energies with the inclusion of the isoscalar pairing at zero and finite temperatures.
Using the SkM* and DD-ME2 functionals without the isoscalar pairing, it is seen that the centroid energies decrease rapidly for T$>$ 0.5 MeV. By including the isoscalar pairing for the SkM* and DD-ME2 functionals, the centroid energies are slightly lowered with increasing temperature. As explained in Sec. \ref{1}, these changes in the centroid energies of the GTRs are related to the competition between the temperature and isoscalar pairing. The temperature leads to decrease of the GTR energies, but at the same time, it reduces the role of the attractive
isoscalar pairing. Therefore, the centroid energies slightly decrease under the influence of the increasing temperature and decreasing isoscalar pairing effects.

\begin{figure}[!ht]
  \begin{center}
\includegraphics[width=1\linewidth,clip=true]{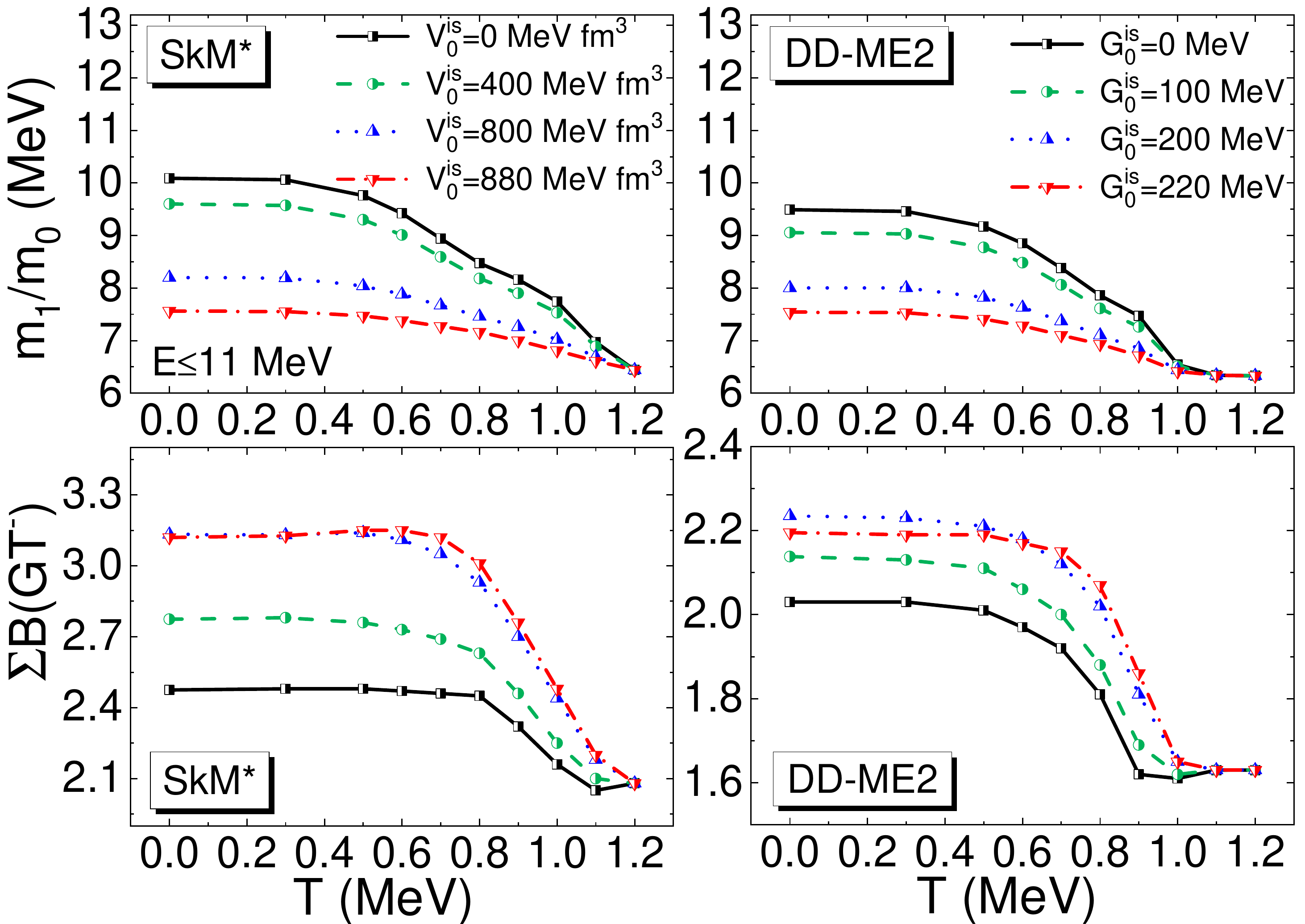}
  \end{center}
 \caption{The same as in Fig. \ref{422}, but for $^{46}$Ti.} 
  \label{466}
\end{figure}

As mentioned above, the behavior of the low-energy excitations under the influence of the temperature and isoscalar pairing is also important for the modeling of astrophysically relevant processes, such as $\beta$-decays, neutrino-nucleus interaction, etc. For a more quantitative analysis, the centroid energies ($m_1/m_0$) and the total sums of the GT$^{-}$ strengths are displayed in Fig. \ref{466} for the excited states below 11 MeV at finite temperatures. The calculations are performed using the SkM* (left panels) and DD-ME2 (right panels) functionals with various isoscalar pairing strengths. It is seen that the behavior of the low-energy states is similar in $^{42}$Ca and $^{46}$Ti.
Without the isoscalar pairing, one can observe that the centroid energies in the low-energy region change rapidly for T$>$0.5 MeV. On the other hand, the centroid energies decrease slowly in comparison to the results without the isoscalar pairing when large pairing strengths are included in the calculations. The behavior of the total low-energy strength is also displayed in the lower panels of Fig. \ref{466}. With and without isoscalar pairing, the total strength remains almost constant up to T=0.8 MeV. At temperatures above, it decreases rapidly in each case. Without the isoscalar pairing, the lowering of the total strength and centroid energy is related to the decrease in the isovector pairing effect and the residual interaction with increasing temperature. In the presence of the isoscalar pairing in the residual interaction, the total strength and centroid energies are affected by increasing temperature due to the decreasing effect of the isoscalar pairing as well as the isovector pairing. Hence the centroid energies slightly decrease, whereas the total strength is rapidly lowered close to the critical temperature. In order to explain these changes in the low-energy strength, the quasiparticle configurations and their contributions to the norm of the states are displayed in Table \ref{tableqq} for the SkM* functional using the largest value of the isoscalar pairing strength at finite temperatures. At zero temperature, the strength of the low-energy peak increases due to the contribution of the ($l=l',j=j'\pm$1) configurations in addition to the ($l=l',j=j'$) ones. By increasing temperature, the contribution of the ($l=l',j=j'\pm$1) configurations slightly change up to T=0.9 MeV, and the low-energy strength slightly decreases. For T$>$0.9 MeV, the contributions of the ($l=l',j=j'\pm$1) configurations also start to decrease, and the lowering of the total strength becomes more pronounced. Above the critical temperature (T=1.2 MeV), the ($l=l',j=j'\pm$1) configurations do not contribute to the low-energy peak due to the disappearance of the pairing correlations, and the low-energy peak is composed of the ($l=l',j=j'$) configurations.

\begin{figure*}[!htb]
  \begin{center}
\includegraphics[width=1\linewidth,clip=true]{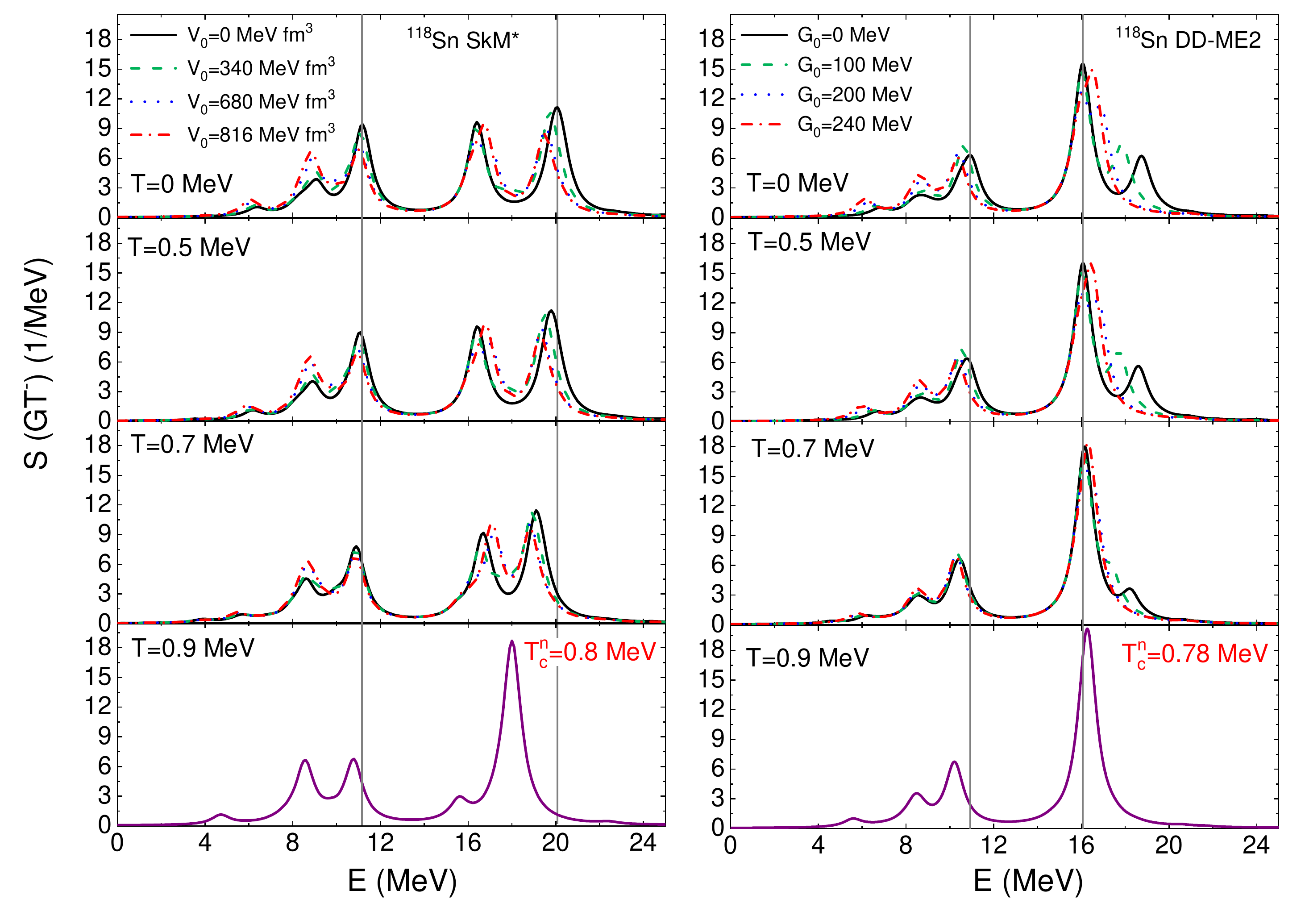}
  \end{center}
 \caption{The same as in Fig. \ref{42}, but for $^{118}$Sn.} 
  \label{118}
\end{figure*}

\begin{table*}
\caption{The quasiparticle configurations that give the major contribution to the selected low-lying GT$^{-}$ states in $^{118}$Sn. The calculations are performed using the SkM* and DD-ME2 interactions with and without isoscalar pairing. The configuration energies and their contribution to the norm of the state (in percentage) [see Eqs. (\ref{Aab}) and (\ref{aaa})] are given.} \label{tableq}
\begin{center}
\begin{tabular}{@{\extracolsep{4pt}}ccccccccc@{}}\hline\hline \\ [-1.5ex]
 && \multicolumn{2}{c}{SkM*} & &\multicolumn{2}{c}{DD-ME2}  \\ \cline{3-4} \cline{5-7} \\[-1.ex]
 &T=0 MeV       &$V_{0}^{is}=0$ MeV fm$^{3}$ & $V_{0}^{is}=816$ MeV fm$^{3}$&&  $G_{0}^{is}=0$ MeV & $G_{0}^{is}=240$ MeV\\ [1.ex] \cline{3-3} \cline{4-4}  \cline{6-6} \cline{7-7} \\ [-1.5ex] 
 &Configurations& E=11.18 MeV & E=11.0 MeV&  & E=10.97 MeV& E=10.38 MeV\\ [1.ex] 
\hline
 & ($\pi 2d_{3/2},\nu 2d_{5/2}$)   & 68.52    & 92.64    &  &22.18    & 83.11 \\
 & ($\pi 1h_{11/2},\nu 1h_{11/2}$) & 18.27    &          &  &57.96    &  \\
 & ($\pi 2d_{5/2},\nu 2d_{5/2}$)   & 3.53     & 1.90     &  &3.80     & 5.52 \\
 & ($\pi 1g_{7/2},\nu 1g_{9/2}$)   & 3.23     & 2.65     &  &6.05     & 4.61 \\
 & ($\pi 3s_{1/2},\nu 3s_{1/2}$)   & 2.33     & 1.34     &  &3.61     & 2.05 \\ 
\hline\hline
\end{tabular}
\end{center}
\vspace{-7mm}
\end{table*}
\begin{table*}
\caption{The same as in Table \ref{tableqq}, but for $^{118}$Sn. The isoscalar pairing is taken as $V_{0}^{is}=816$ MeV fm$^{3}$.} \label{tableqqq}
\begin{center}
\begin{tabular}{@{\extracolsep{4pt}}cccccccc@{}}
\hline\hline \\ [-1.5ex]
 &SkM*          &T=0 MeV & T=0.5 MeV & T=0.7 MeV & T=0.9 MeV\\ [1.ex] \cline{3-3} \cline{4-4}\cline{5-5} \cline{6-6} \cline{7-7} \\ [-1.5ex] 
 && E=8.86 MeV &E=8.77 MeV&E=8.60 MeV   &E=8.56 MeV\\ 
 &Configurations& B(GT$^{-}$)=8.14 & B(GT$^{-}$)=8.27& B(GT$^{-}$)=6.90    & B(GT$^{-}$)=9.41 \\ 
\hline
 & ($\pi 2d_{3/2},\nu 2d_{3/2}$)   & 4.46    & 5.71   & 8.28   &2.76      \\
 & ($\pi 3s_{1/2},\nu 3s_{1/2}$)   & 44.92   & 36.57  & 26.08  & 14.65  \\
 & ($\pi 1g_{7/2},\nu 1g_{7/2}$)   & 7.14    & 9.77   & 14.68  & 5.91   \\
 & ($\pi 2d_{5/2},\nu 2d_{5/2}$)   & 24.28   & 28.0   & 29.40  & 39.64  \\
 & ($\pi 1g_{7/2},\nu 2d_{5/2}$)   & 3.20    & 3.68   & 3.71   & 7.30   \\
 & ($\pi 2d_{3/2},\nu 3s_{1/2}$)   & 9.63    & 9.54   & 8.32   & 5.18     \\
 & ($\pi 2d_{3/2},\nu 2d_{5/2}$)   & 3.85    & 4.07   & 3.88   &  4.82     \\
 & ($\pi 1h_{11/2},\nu 1h_{11/2}$) & 1.15    &        & 3.38   & 16.11      \\
\hline\hline
\end{tabular}
\end{center}
\vspace{-7mm}
\end{table*}

\subsection{\texorpdfstring{$^{118}$Sn} {} nucleus}
\label{4}
Finally, we discuss the behavior of the GT$^{-}$ excitations in $^{118}$Sn under the influence of temperature and pairing. In this way, we can test the validity of our findings in Secs. \ref{1} and \ref{3} using a heavier nucleus. For $^{118}$Sn, the isovector pairing strengths for neutrons and protons are taken as $V_{0}^{iv}$=680 MeV fm$^{3}$ and $G_{0}^{iv}$=26 MeV/A, and the critical temperature values for neutrons are obtained at T$^{n}_{c}$=0.8 and 0.78 MeV using the SkM* and DD-ME2 functionals, respectively. At zero temperature, $^{118}$Sn has proton shell closure at Z=50, and the neutron states are partially occupied due to the isovector pairing. By increasing the temperature, the occupation probabilities of proton $2p_{1/2}$ and $1g_{9/2}$ states start to decrease, while $2d_{5/2}$, $1g_{7/2}$, $3s_{1/2}$, and $2d_{3/2}$ states become populated. 

In Fig. \ref{118}, the GT$^{-}$ strength is displayed by varying the value of the isoscalar pairing strength at finite temperatures. Similar to the findings in Secs. \ref{1} and \ref{3}, the low-energy states start to shift slightly downwards and the strength increases. In addition, the strength and the energy difference between the GTR peaks slightly decrease with increasing isoscalar pairing strength at zero temperature (see the top most panels in Fig. \ref{118}). However, the effect of the isoscalar pairing in $^{118}$Sn is found to be smaller compared to $^{42}$Ca and $^{46}$Ti. Using the SkM* functional without the isoscalar pairing, the GTR peaks are obtained at 16.42, 19.90 and 20.20 MeV. While the first peak is dominated by the ($\pi 1g_{7/2},\nu 1g_{9/2}$) configuration, the latter two are mainly composed of the ($\pi 1h_{9/2},\nu 1h_{11/2}$) and ($\pi 2h_{9/2},\nu 1h_{11/2}$) configurations. By increasing the isoscalar pairing strength, the residual particle-particle matrix elements of the ($\pi 1h_{9/2},\nu 1h_{11/2}$) and ($\pi 1h_{11/2},\nu 1h_{9/2}$) configurations are affected more as compared to the ($\pi 1g_{7/2},\nu 1g_{9/2}$) due to the $u$ and $v$ factors of the related quasi-particle states. Therefore, the ($\pi 1h_{11/2},\nu 1h_{9/2}$) configuration is admixed in the first GTR peak wave function due to the attractive nature of the residual isoscalar pairing, and its contribution to the first GTR peak increases. Eventually, the first GTR peak is composed of the ($\pi 1g_{7/2},\nu 1g_{9/2}$) and ($\pi 1h_{11/2},\nu 1h_{9/2}$) configurations, and its excitation energy slightly increases due to the high unperturbed two q.p. energy of the latter one. For instance, the first GTR peak is obtained at 16.77 MeV using the SkM* functional for the largest value of the isoscalar pairing strength. The energy of the second GTR peak also slightly decreases due to the impact of the attractive isoscalar pairing on the ($\pi 1h_{9/2},\nu 1h_{11/2}$) configuration, and it is obtained at 19.48 MeV. Similar results are also obtained using the DD-ME2 functional, whereas the GTR peaks do not have comparable strengths. As mentioned in Sec. \ref{1}, the differences in the structure of the GTR states are related to the differences in the predicted shell structure of nuclei due to the effective interactions used in the calculations. For instance, the first and second GTR peaks are obtained at 16.06 and 18.71 MeV in the case of no isoscalar pairing, and the former one carries most of the GTR strength. For the largest isoscalar pairing strength, the GTR peaks are obtained at 15.91 and 16.56 MeV, and the latter one has more strength. In the low-energy region, it is seen that the strength of the excited states around 8.5 MeV is enhanced due to the isoscalar pairing, whereas the strengths of the states around 11.0 MeV slightly decrease. For the low-energy region, the main configurations for the selected excited states can be followed in Table \ref{tableq}. One can observe that the isoscalar pairing acts mainly to the two q.p. configurations on the ($l=l',j=j'\pm$1) and in this way it modifies the contributions to the excited states. For the selected excited states, the isoscalar pairing increases the contribution of the ($\pi 2d_{3/2},\nu 2d_{5/2}$), while the contribution of the ($\pi 1h_{11/2},\nu 1h_{11/2}$) configuration is removed for both the SkM* and DD-ME2 functionals. Therefore, the collectivity and the strength of the selected excited states decrease. The GT$^{-}$ excitations in $^{118}$Sn have already been studied using the relativistic \cite{PhysRevC.69.054303} and nonrelativistic \cite{PhysRevC.76.044307,PhysRevC.95.044301} functionals at zero temperature, and similar results were obtained.

We continue our analysis by varying the temperature in the range from T=0 toward T=0.9 MeV, with and without isoscalar pairing in the residual FT-PNQRPA interaction (see Fig. \ref{118}). Compared to the results for $^{42}$Ca and $^{46}$Ti, the $^{118}$Sn nucleus is weakly sensitive on the temperature and pairing effects. Using the SkM* functional with and without isoscalar pairing, it is seen that the first GTR peak only slightly starts to move upwards, while the second GTR peak shifts downwards with increasing temperature. Without the isoscalar pairing, the increase (decrease) in the first (second) GTR peak is related to the changes in the single(quasi)-particle states and weakening of the residual interaction with increasing temperature. As explained above, the unperturbed energies of the two q.p. configurations are lowered due to the reduction of the isovector pairing correlations with increasing temperature. Compared to the ($\pi 1g_{7/2},\nu 1g_{9/2}$) configuration, the unperturbed energy of the ($\pi 1h_{9/2},\nu 1h_{11/2}$) configuration is influenced more and decreases more with the temperature. Therefore, the ($\pi 1h_{9/2},\nu 1h_{11/2}$) configuration starts to mix further with the ($\pi 1g_{7/2},\nu 1g_{9/2}$) and slightly increases the energy of the first GTR peak with increasing temperature. For instance, at T=0.7 MeV, the first GTR peak is obtained at 16.64 for the SkM* functional. As mentioned above, the second GTR peak is mainly composed of the ($\pi 1h_{9/2},\nu 1h_{11/2}$) configuration at zero temperature. By increasing temperature, the contribution of this configuration for the second GTR peak decreases, and other configurations also start to contribute to this excited state. In addition, the excitation energy of the second GTR peak decreases with increasing temperature, as expected. For instance, the second GTR peak is obtained at 19.07 MeV using the SkM* functional at T=0.7 MeV. Similar results are also obtained using the DD-ME2 functional without the isoscalar pairing, whereas the GTR is mainly accumulated in a single peak and the centroid energy slightly increases with increasing temperature.
For the calculations including the isoscalar pairing, the energies of the GTRs change slowly due to the interplay of temperature and isoscalar pairing effects. 
Including the isoscalar pairing, the first (second) GTR peak is obtained at slightly higher (lower) energies for the SkM* functional, compared to the results without the isoscalar pairing at finite temperatures. For the DD-ME2 functional, the centroid energy of the GTR almost do not change for the calculations using large isoscalar pairing strength below the critical temperature. Above the critical temperature, i.e., at T=0.9 MeV, the pairing correlations vanish and a single peak is obtained in the GTR region at 18.0 (16.27) MeV for the SkM* (DD-ME2) functional. This peak is also composed of the ($\pi 1g_{7/2},\nu 1g_{9/2}$) and ($\pi 1h_{9/2},\nu 1h_{11/2}$) configurations with comparable contributions.

It is also seen that the low-energy states do not display strong sensitivity to the changes in the isoscalar pairing strength or temperature. Without the isoscalar pairing, the low-energy states start to shift downward with increasing temperature, whereas these states are more stable against temperature effects with the inclusion of the large isoscalar pairing strength in the calculations. In case without the isoscalar pairing, the strength of the peaks around 8.5 MeV become more pronounced, while the total strength below 12 MeV remains almost constant with increasing temperature. Using the large isoscalar pairing strength in the calculations, the energy and strength of the excited states around 8.5 and 10.5 MeV are only little affected by the temperature due to the interplay between the isoscalar pairing and temperature effects. In Table \ref{tableqqq}, the configurations of the selected excited states are given at finite temperatures for the SkM* functional. The selected excited states are mainly formed with the ($l=l',j=j'$) configurations and display collectivity due to the contribution of the several q.p. configurations with comparable weights both at zero and finite temperatures. By increasing temperature, the contribution of the configurations to the given excited state is changed due to the changes in the pairing properties with increasing temperature. In contradistinction to the findings for $^{42}$Ca and $^{46}$Ti, the low-energy states display collectivity and the total strength below 12 MeV slightly decrease for $^{118}$Sn. Considering our findings, the low or high sensitivity of the GT$^{-}$ strength to the isoscalar pairing and temperature effects is related to the details of the single-particle spectra of nuclei.

\section{conclusions}
\label{finito}
In this work, the self-consistent finite temperature PNQRPA is developed for the first time to study the finite temperature and pairing effects on the Gamow-Teller response of nuclei. In order to explore the model dependence of the GT$^{-}$ transitions in open-shell nuclei, two independent FT-PNQRPA frameworks have been established, based on the relativistic (DD-ME2) and nonrelativistic Skyrme (SkM*) nuclear energy density functionals. In addition, both the isoscalar and isovector pairing
correlations are taken into account, which are crucial
for the proper description of the Gamow-Teller states
in open-shell nuclei. Using the nonrelativistic
and relativistic FT-PNQRPA, the effects of the pairing
and temperature are studied in the case of the Gamow-Teller excitations in $^{42}$Ca, $^{46}$Ti, and $^{118}$Sn nuclei at low temperatures.
and temperature are studied in the case of the Gamow-Teller excitations in $^{42}$Ca, $^{46}$Ti, and $^{118}$Sn nuclei at low temperatures.

With the increase of isoscalar pairing strength
at zero temperature, the GTRs start to shift downwards and the transition strength slightly decreases,
while the low-lying states move downward in energy and their strengths increase. Without the isoscalar pairing, the low-energy states are formed mainly from the ($l=l', j=j'$) transitions. With the inclusion of the isoscalar pairing, ($l=l', j=j'\pm$1) transitions start to contribute to the excited states in a more coherent way and their strength becomes larger. Although the effect of the isoscalar pairing is comparable in all nuclei considered in this work, it is shown that the GT$^{-}$ states are more sensitive to its effects in $^{42}$Ca and $^{46}$Ti because the results are more sensitive on the details of the single-particle spectra. Our findings are consistent with previous studies addressing the isoscalar pairing in 
Refs. \cite{PhysRevC.69.054303,PhysRevC.76.044307,PhysRevC.95.044301,PhysRevC.90.054335}.

By increasing temperature, the GT$^{-}$ states start to shift downwards, and additional excited states are obtained with and without isoscalar pairing in all considered nuclei. In the cases without the isoscalar pairing, the changes in the GT$^{-}$ states are caused by the decrease of the isovector pairing effects and the softening of the repulsive $ph$ interaction due to the temperature factors with increasing temperature. In addition, these changes become more apparent close to the critical temperatures due to the rapid decrease of the isovector pairing properties of nuclei. 
Including isoscalar pairing in the calculations, the predictions for the GT$^{-}$ strengths and excitation energies are rather different compared to the results without the isoscalar pairing at finite temperatures. With the inclusion of the isoscalar pairing, the decrease in the excitation energies depends on the competition between the temperature and isoscalar pairing effects. We find that the temperature reduces the impact of the attractive isoscalar pairing and the properties of the excited states depend on the interplay between the increasing effect of the temperature and decreasing impact of the isoscalar pairing. Therefore, in a complete calculation the excited states slowly shift downwards in energy compared to the results without the isoscalar pairing with increasing temperature. The strengths of the GT$^{-}$ states are also impacted: the temperature reduces the contributions of the ($l=l', j=j'\pm$1) transitions to the excited states by decreasing the impact of the isoscalar pairing. Hence the low-energy strength is reduced and the GTR strength is enhanced with increasing temperature. We also find that these effects are more pronounced for the isoscalar pairing sensitive $^{42}$Ca and $^{46}$Ti nuclei. Increasing the temperature and isoscalar pairing slightly impacts the strength and excitation energies of $^{118}$Sn.

The development of the FT-PNQRPA framework is relevant not only for a detailed 
understanding of the excitation phenomena in nuclei at finite temperature, but also for
consistent and universal modeling of the weak-interaction processes in stellar environments, such as electron capture, $\beta$-decays, and neutrino-nucleus reactions \cite{RevModPhys.75.819,JANKA200738}. Improved description of weak-interaction processes, in a framework that includes both the pairing and finite temperature effects, could have important consequences for complete understanding of the core-collapse supernova explosion mechanism as well as nucleosynthesis. As a first step in a forthcoming study, the FT-PNQRPA will be employed in modeling the electron capture rates for implementation in core-collapse supernova simulations.

\begin{acknowledgments}
E. Y. acknowledges financial support from the Scientific
and Technological Research Council of Turkey (T\"{U}B\.{I}TAK) BIDEB-2219 Postdoctoral Research program. G. C. acknowledges funding from the European Union's Horizon 2020 research
and innovation program under Grant No. 654002. Y. F. N. acknowledges the support from the Fundamental Research Funds for the Central Universities under Grant No. Lzujbky-2019-11. This work is supported by the Croatian Science Foundation under the project Structure and Dynamics of Exotic Femtosystems (IP-2014-09-9159) and by the QuantiXLie Centre of Excellence, a project co financed by the Croatian Government and European Union through the European Regional Development Fund, the Competitiveness and Cohesion Operational Programme (KK.01.1.1.01). This article is based upon work from the ``ChETEC'' COST Action (CA16117), supported by COST (European Cooperation in Science and Technology).
\end{acknowledgments}

\appendix
\section{COMPARISON OF THE RESULTS WITH EXPERIMENT}
\label{appendix}
In this appendix, we provide details for the calculation of the excited state energies with respect to the daughter nucleus.
In the proton-neutron RPA (pn-RPA) approach, we construct a basis that includes both proton particle-neutron hole ($\Delta T_z=-1$) and neutron particle-proton hole ($\Delta T_z=+1$) configurations based on the target (parent) nucleus ground-state. We will focus first on equations for the $\Delta T_{z}=-1$ channel. The energy of the proton-neutron configuration with respect to the target nucleus should be  
 
\begin{equation} 
E_{conf-t}=(m_pc^2+\varepsilon_p)-(m_nc^2+\varepsilon_n).
\label{10}
\end{equation}

However, in the pn-RPA calculations, we only calculate 

\begin{equation} 
E_{conf-RPA}=\varepsilon_p-\varepsilon_n,
\label{20}
\end{equation}

and the proton-neutron mass difference is missing. Therefore, the actual excitation energy $E^*_{RPA}$ is given by

\begin{equation}
\begin{split}
E^*_{RPA}&=E^*_{t}+(m_n-m_p)c^2 \\
       &=E^*_{t}+\Delta np,
\end{split}
\label{30}
\end{equation}
where $E_{RPA}^{*}$ is the calculated excited state energy using the RPA, $E^*_{t}$ is the excitation energy with respect to the target nucleus ground-state, and $\Delta np$ is the  mass difference between neutron and proton.

\begin{figure}[!htb]
  \begin{center}
\includegraphics[width=1\linewidth,clip=true]{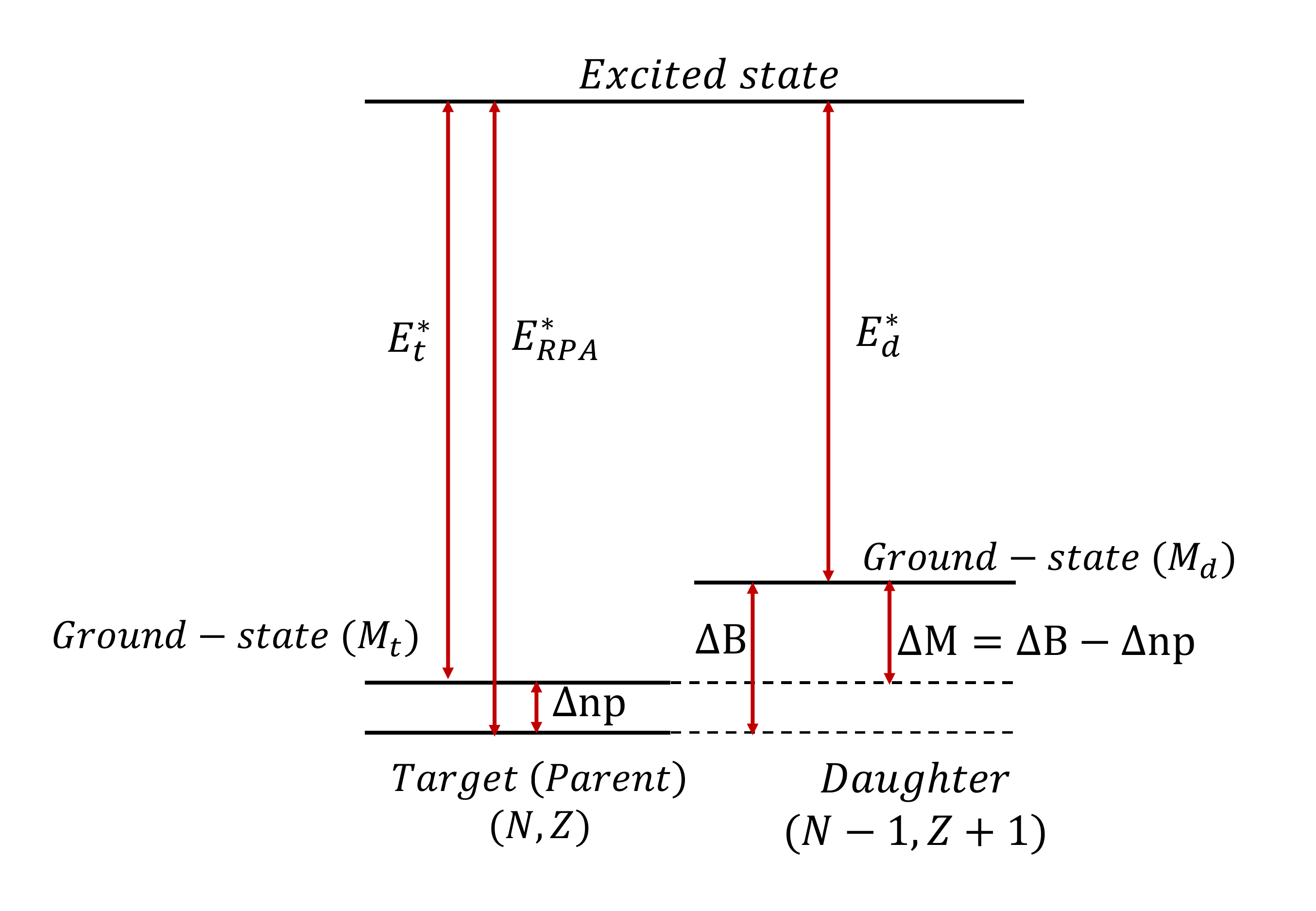}
\includegraphics[width=1\linewidth,clip=true]{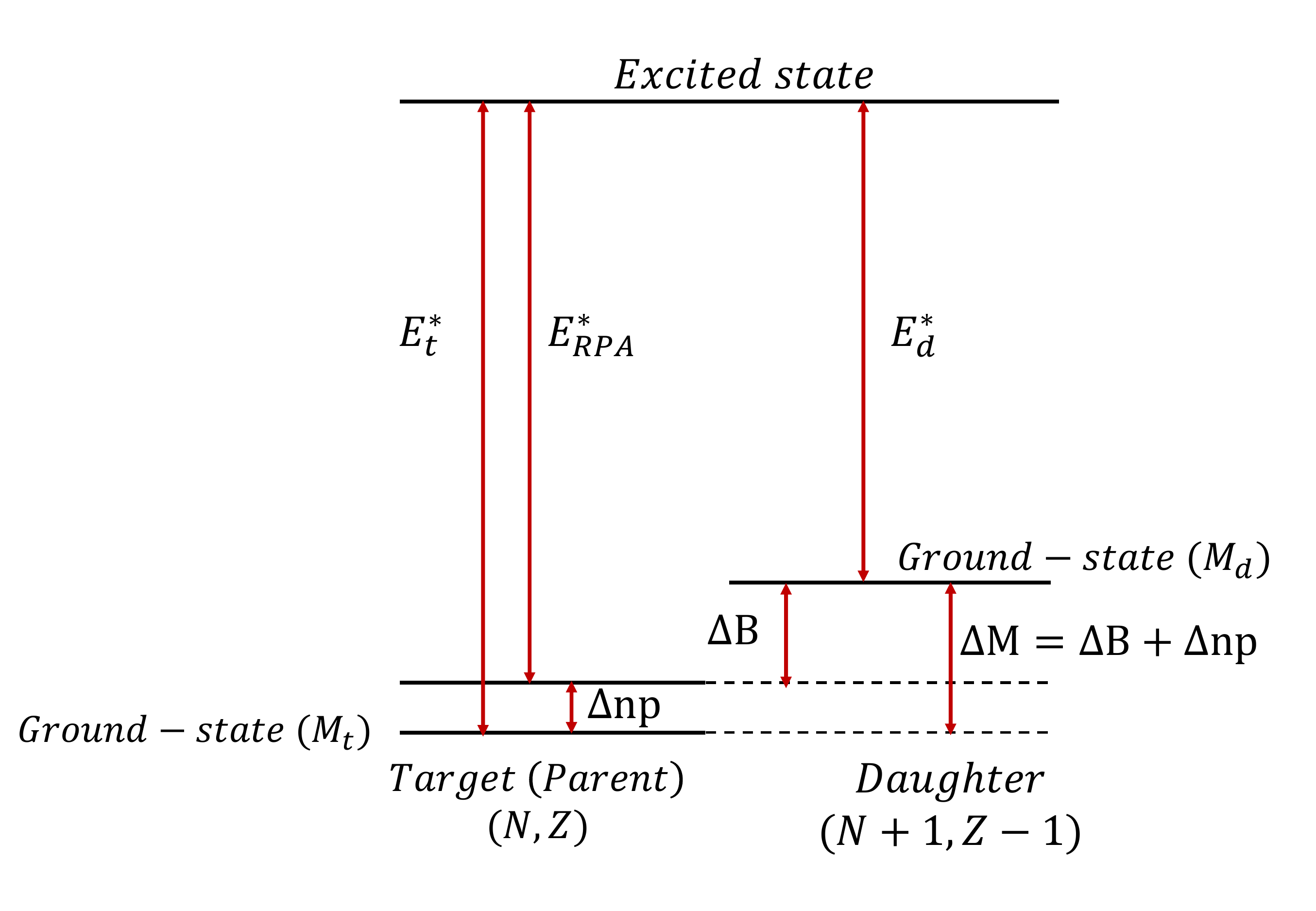}
  \end{center}
 \caption{Schematic representation of the charge-exchange excitation in the $\Delta T_{z}=-1$ (upper panel) and  $\Delta T_{z}=1$ (lower panel) channels. 
$E_{RPA}^{*}$ denotes the RPA excitation energy, $E_{t}^{*}$ and $E_{d}^{*}$ 
are the excitation energies with respect to the target and
the daughter ground-state, respectively. The mass and binding energy differences between the daughter and target nucleus are given by $\Delta M= M_d-M_t$ and $\Delta B= B_t-B_d$, respectively. $\Delta np$ represents the neutron-proton mass difference.
} 
  \label{fig}
\end{figure}

Schematic representations of the charge-exchange excitation for the $\Delta T_{z}=\pm 1$ channels are presented in Fig. \ref{fig}. From Fig. \ref{fig}, it is clearly seen that $E_{t}^{*}$ for the $\Delta T_{z}=-1$ channel can also be written as

\begin{equation} 
E_{t}^{*}=E_{d}^{*}+ \Delta M,
\label{40}
\end{equation}

where $E_{d}^{*}$ is the excitation energy with respect to the daughter nucleus ground-state, and the mass difference ($\Delta M$) between the daughter and target nucleus is calculated as

\begin{equation} 
\begin{split}
\Delta M&= M_d-M_t \\
      &=\left[(N-1)m_n+(Z+1)m_p-B_d\right]-\left[N m_n+Zm_p-B_t\right] \\
			&=-m_n+m_p+B_t-B_d  \\
			&= B_t-B_d-\Delta np=\Delta B_{t-d}-\Delta np.
\end{split}
\label{50}
\end{equation}
Here, $B_{d(t)}$ is the binding energy of daughter (target) nucleus. Inserting Eq. (\ref{50}) into Eq. (\ref{40}), excitation energy with respect to the target nucleus ground-state can be written as
 
\begin{equation} 
E_{t}^{*}=E_{d}^{*}+\Delta  B_{t-d}-\Delta np.
\label{60}
\end{equation}

Finally, using Eqs. (\ref{30}) and (\ref{60}), the excited state energies with respect to the daughter nucleus is obtained as

\begin{equation} 
E_{d}^{*}=E_{RPA}^{*}-\Delta  B_{t-d}.
\label{70}
\end{equation}

It should be noted that the Eq. (\ref{70}) is also valid for the $\Delta T_{z}=1$ channel of the excitation. For the $\Delta T_{z}=1$ channel, we use the difference between the proton and neutron masses ($\Delta pn$) in the derivation of the equations instead of the difference between the neutron and proton masses ($\Delta np$).

In the pn-QRPA approach, we diagonalize $H' = H - \lambda N$ and not $H$. Therefore, we also need to include Fermi energies ($\lambda_{p(n)}$), and the actual
excitation energy for the $\Delta T_{z}= \pm 1$ channel of the pn-QRPA excitation is given by 
\begin{equation} 
E_{QRPA}^{*}=E_{t}^{*}\mp(m_n-m_p)c^{2} \mp (\lambda_n-\lambda_p).
\label{13}
\end{equation}

Similar to Eq. (\ref{70}) and for the $\Delta T_{z}= \pm 1$ channel of the pn-QRPA excitation, the excited state energies with respect to the daughter nucleus can be written as 

\begin{equation} 
E_{d}^{*}=E_{QRPA}^{*}-\Delta  B_{t-d}\pm (\lambda_n-\lambda_p).
\label{80}
\end{equation}

We should also point out that the $\Delta T_{z}=1$ and $\Delta T_{z}=-1$ channels are coupled in QRPA. Although the channels are coupled, QRPA solutions should correspond to either one channel or another, i.e., the solutions have good $T_z$.
To compare with the experimental values of the charge-exchange resonances, which are usually provided in the final, or daughter systems, we transform
the (Q)RPA solution into the corresponding value with respect to the final ground-state by using experimental binding energies, according to Eqs. (\ref{70}) and (\ref{80}).

\bibliography{apssamp}

\providecommand{\noopsort}[1]{}\providecommand{\singleletter}[1]{#1}%
\begin{thebibliography}{68}%
\makeatletter
\providecommand \@ifxundefined [1]{%
 \@ifx{#1\undefined}
}%
\providecommand \@ifnum [1]{%
 \ifnum #1\expandafter \@firstoftwo
 \else \expandafter \@secondoftwo
 \fi
}%
\providecommand \@ifx [1]{%
 \ifx #1\expandafter \@firstoftwo
 \else \expandafter \@secondoftwo
 \fi
}%
\providecommand \natexlab [1]{#1}%
\providecommand \enquote  [1]{``#1''}%
\providecommand \bibnamefont  [1]{#1}%
\providecommand \bibfnamefont [1]{#1}%
\providecommand \citenamefont [1]{#1}%
\providecommand \href@noop [0]{\@secondoftwo}%
\providecommand \href [0]{\begingroup \@sanitize@url \@href}%
\providecommand \@href[1]{\@@startlink{#1}\@@href}%
\providecommand \@@href[1]{\endgroup#1\@@endlink}%
\providecommand \@sanitize@url [0]{\catcode `\\12\catcode `\$12\catcode
  `\&12\catcode `\#12\catcode `\^12\catcode `\_12\catcode `\%12\relax}%
\providecommand \@@startlink[1]{}%
\providecommand \@@endlink[0]{}%
\providecommand \url  [0]{\begingroup\@sanitize@url \@url }%
\providecommand \@url [1]{\endgroup\@href {#1}{\urlprefix }}%
\providecommand \urlprefix  [0]{URL }%
\providecommand \Eprint [0]{\href }%
\providecommand \doibase [0]{http://dx.doi.org/}%
\providecommand \selectlanguage [0]{\@gobble}%
\providecommand \bibinfo  [0]{\@secondoftwo}%
\providecommand \bibfield  [0]{\@secondoftwo}%
\providecommand \translation [1]{[#1]}%
\providecommand \BibitemOpen [0]{}%
\providecommand \bibitemStop [0]{}%
\providecommand \bibitemNoStop [0]{.\EOS\space}%
\providecommand \EOS [0]{\spacefactor3000\relax}%
\providecommand \BibitemShut  [1]{\csname bibitem#1\endcsname}%
\let\auto@bib@innerbib\@empty
\bibitem [{\citenamefont {Osterfeld}(1992)}]{RevModPhys.64.491}%
  \BibitemOpen
  \bibfield  {author} {\bibinfo {author} {\bibfnamefont {F.}~\bibnamefont
  {Osterfeld}},\ }\href {\doibase 10.1103/RevModPhys.64.491} {\bibfield
  {journal} {\bibinfo  {journal} {Rev. Mod. Phys.}\ }\textbf {\bibinfo {volume}
  {64}},\ \bibinfo {pages} {491} (\bibinfo {year} {1992})}\BibitemShut
  {NoStop}%
\bibitem [{\citenamefont {Ichimura}\ \emph {et~al.}(2006)\citenamefont
  {Ichimura}, \citenamefont {Sakai},\ and\ \citenamefont
  {Wakasa}}]{ICHIMURA2006446}%
  \BibitemOpen
  \bibfield  {author} {\bibinfo {author} {\bibfnamefont {M.}~\bibnamefont
  {Ichimura}}, \bibinfo {author} {\bibfnamefont {H.}~\bibnamefont {Sakai}}, \
  and\ \bibinfo {author} {\bibfnamefont {T.}~\bibnamefont {Wakasa}},\ }\href
  {\doibase 10.1016/j.ppnp.2005.09.001} {\bibfield  {journal} {\bibinfo
  {journal} {Progress in Particle and Nuclear Physics}\ }\textbf {\bibinfo
  {volume} {56}},\ \bibinfo {pages} {446 } (\bibinfo {year}
  {2006})}\BibitemShut {NoStop}%
\bibitem [{\citenamefont {Fujita}\ \emph {et~al.}(2011)\citenamefont {Fujita},
  \citenamefont {Rubio},\ and\ \citenamefont {Gelletly}}]{FUJITA2011549}%
  \BibitemOpen
  \bibfield  {author} {\bibinfo {author} {\bibfnamefont {Y.}~\bibnamefont
  {Fujita}}, \bibinfo {author} {\bibfnamefont {B.}~\bibnamefont {Rubio}}, \
  and\ \bibinfo {author} {\bibfnamefont {W.}~\bibnamefont {Gelletly}},\ }\href
  {\doibase 10.1016/j.ppnp.2011.01.056} {\bibfield  {journal} {\bibinfo
  {journal} {Progress in Particle and Nuclear Physics}\ }\textbf {\bibinfo
  {volume} {66}},\ \bibinfo {pages} {549 } (\bibinfo {year}
  {2011})}\BibitemShut {NoStop}%
\bibitem [{\citenamefont {Langanke}\ and\ \citenamefont
  {Mart\'{\i}nez-Pinedo}(2003)}]{RevModPhys.75.819}%
  \BibitemOpen
  \bibfield  {author} {\bibinfo {author} {\bibfnamefont {K.}~\bibnamefont
  {Langanke}}\ and\ \bibinfo {author} {\bibfnamefont {G.}~\bibnamefont
  {Mart\'{\i}nez-Pinedo}},\ }\href {\doibase 10.1103/RevModPhys.75.819}
  {\bibfield  {journal} {\bibinfo  {journal} {Rev. Mod. Phys.}\ }\textbf
  {\bibinfo {volume} {75}},\ \bibinfo {pages} {819} (\bibinfo {year}
  {2003})}\BibitemShut {NoStop}%
\bibitem [{\citenamefont {Janka}\ \emph {et~al.}(2007)\citenamefont {Janka},
  \citenamefont {Langanke}, \citenamefont {Marek}, \citenamefont
  {Mart\'{\i}nez-Pinedo},\ and\ \citenamefont {M\"{u}ller}}]{JANKA200738}%
  \BibitemOpen
  \bibfield  {author} {\bibinfo {author} {\bibfnamefont {H.~T.}\ \bibnamefont
  {Janka}}, \bibinfo {author} {\bibfnamefont {K.}~\bibnamefont {Langanke}},
  \bibinfo {author} {\bibfnamefont {A.}~\bibnamefont {Marek}}, \bibinfo
  {author} {\bibfnamefont {G.}~\bibnamefont {Mart\'{\i}nez-Pinedo}}, \ and\
  \bibinfo {author} {\bibfnamefont {B.}~\bibnamefont {M\"{u}ller}},\ }\href
  {\doibase 10.1016/j.physrep.2007.02.002} {\bibfield  {journal} {\bibinfo
  {journal} {Physics Reports}\ }\textbf {\bibinfo {volume} {442}},\ \bibinfo
  {pages} {38} (\bibinfo {year} {2007})}\BibitemShut {NoStop}%
\bibitem [{\citenamefont {Caurier}\ \emph {et~al.}(1999)\citenamefont
  {Caurier}, \citenamefont {Langanke}, \citenamefont {Mart\'{\i}nez-Pinedo},\
  and\ \citenamefont {Nowacki}}]{CAURIER1999439}%
  \BibitemOpen
  \bibfield  {author} {\bibinfo {author} {\bibfnamefont {E.}~\bibnamefont
  {Caurier}}, \bibinfo {author} {\bibfnamefont {K.}~\bibnamefont {Langanke}},
  \bibinfo {author} {\bibfnamefont {G.}~\bibnamefont {Mart\'{\i}nez-Pinedo}}, \
  and\ \bibinfo {author} {\bibfnamefont {F.}~\bibnamefont {Nowacki}},\ }\href
  {\doibase 10.1016/S0375-9474(99)00240-7} {\bibfield  {journal} {\bibinfo
  {journal} {Nuclear Physics A}\ }\textbf {\bibinfo {volume} {653}},\ \bibinfo
  {pages} {439 } (\bibinfo {year} {1999})}\BibitemShut {NoStop}%
\bibitem [{\citenamefont {Yoshida}\ \emph {et~al.}(2018)\citenamefont
  {Yoshida}, \citenamefont {Utsuno}, \citenamefont {Shimizu},\ and\
  \citenamefont {Otsuka}}]{PhysRevC.97.054321}%
  \BibitemOpen
  \bibfield  {author} {\bibinfo {author} {\bibfnamefont {S.}~\bibnamefont
  {Yoshida}}, \bibinfo {author} {\bibfnamefont {Y.}~\bibnamefont {Utsuno}},
  \bibinfo {author} {\bibfnamefont {N.}~\bibnamefont {Shimizu}}, \ and\
  \bibinfo {author} {\bibfnamefont {T.}~\bibnamefont {Otsuka}},\ }\href
  {\doibase 10.1103/PhysRevC.97.054321} {\bibfield  {journal} {\bibinfo
  {journal} {Phys. Rev. C}\ }\textbf {\bibinfo {volume} {97}},\ \bibinfo
  {pages} {054321} (\bibinfo {year} {2018})}\BibitemShut {NoStop}%
\bibitem [{\citenamefont {Saxena}\ \emph {et~al.}(2018)\citenamefont {Saxena},
  \citenamefont {Srivastava},\ and\ \citenamefont
  {Suzuki}}]{PhysRevC.97.024310}%
  \BibitemOpen
  \bibfield  {author} {\bibinfo {author} {\bibfnamefont {A.}~\bibnamefont
  {Saxena}}, \bibinfo {author} {\bibfnamefont {P.~C.}\ \bibnamefont
  {Srivastava}}, \ and\ \bibinfo {author} {\bibfnamefont {T.}~\bibnamefont
  {Suzuki}},\ }\href {\doibase 10.1103/PhysRevC.97.024310} {\bibfield
  {journal} {\bibinfo  {journal} {Phys. Rev. C}\ }\textbf {\bibinfo {volume}
  {97}},\ \bibinfo {pages} {024310} (\bibinfo {year} {2018})}\BibitemShut
  {NoStop}%
\bibitem [{\citenamefont {Paar}\ \emph {et~al.}(2004)\citenamefont {Paar},
  \citenamefont {Nik\ifmmode \check{s}\else \v{s}\fi{}i\ifmmode~\acute{c}\else
  \'{c}\fi{}}, \citenamefont {Vretenar},\ and\ \citenamefont
  {Ring}}]{PhysRevC.69.054303}%
  \BibitemOpen
  \bibfield  {author} {\bibinfo {author} {\bibfnamefont {N.}~\bibnamefont
  {Paar}}, \bibinfo {author} {\bibfnamefont {T.}~\bibnamefont {Nik\ifmmode
  \check{s}\else \v{s}\fi{}i\ifmmode~\acute{c}\else \'{c}\fi{}}}, \bibinfo
  {author} {\bibfnamefont {D.}~\bibnamefont {Vretenar}}, \ and\ \bibinfo
  {author} {\bibfnamefont {P.}~\bibnamefont {Ring}},\ }\href {\doibase
  10.1103/PhysRevC.69.054303} {\bibfield  {journal} {\bibinfo  {journal} {Phys.
  Rev. C}\ }\textbf {\bibinfo {volume} {69}},\ \bibinfo {pages} {054303}
  (\bibinfo {year} {2004})}\BibitemShut {NoStop}%
\bibitem [{\citenamefont {Niu}\ \emph {et~al.}(2017)\citenamefont {Niu},
  \citenamefont {Niu}, \citenamefont {Liang}, \citenamefont {Long},\ and\
  \citenamefont {Meng}}]{PhysRevC.95.044301}%
  \BibitemOpen
  \bibfield  {author} {\bibinfo {author} {\bibfnamefont {Z.~M.}\ \bibnamefont
  {Niu}}, \bibinfo {author} {\bibfnamefont {Y.~F.}\ \bibnamefont {Niu}},
  \bibinfo {author} {\bibfnamefont {H.~Z.}\ \bibnamefont {Liang}}, \bibinfo
  {author} {\bibfnamefont {W.~H.}\ \bibnamefont {Long}}, \ and\ \bibinfo
  {author} {\bibfnamefont {J.}~\bibnamefont {Meng}},\ }\href {\doibase
  10.1103/PhysRevC.95.044301} {\bibfield  {journal} {\bibinfo  {journal} {Phys.
  Rev. C}\ }\textbf {\bibinfo {volume} {95}},\ \bibinfo {pages} {044301}
  (\bibinfo {year} {2017})}\BibitemShut {NoStop}%
\bibitem [{\citenamefont {Liang}\ \emph {et~al.}(2008)\citenamefont {Liang},
  \citenamefont {Van~Giai},\ and\ \citenamefont
  {Meng}}]{PhysRevLett.101.122502}%
  \BibitemOpen
  \bibfield  {author} {\bibinfo {author} {\bibfnamefont {H.}~\bibnamefont
  {Liang}}, \bibinfo {author} {\bibfnamefont {N.}~\bibnamefont {Van~Giai}}, \
  and\ \bibinfo {author} {\bibfnamefont {J.}~\bibnamefont {Meng}},\ }\href
  {\doibase 10.1103/PhysRevLett.101.122502} {\bibfield  {journal} {\bibinfo
  {journal} {Phys. Rev. Lett.}\ }\textbf {\bibinfo {volume} {101}},\ \bibinfo
  {pages} {122502} (\bibinfo {year} {2008})}\BibitemShut {NoStop}%
\bibitem [{\citenamefont {Robin}\ and\ \citenamefont
  {Litvinova}(2018)}]{PhysRevC.98.051301}%
  \BibitemOpen
  \bibfield  {author} {\bibinfo {author} {\bibfnamefont {C.}~\bibnamefont
  {Robin}}\ and\ \bibinfo {author} {\bibfnamefont {E.}~\bibnamefont
  {Litvinova}},\ }\href {\doibase 10.1103/PhysRevC.98.051301} {\bibfield
  {journal} {\bibinfo  {journal} {Phys. Rev. C}\ }\textbf {\bibinfo {volume}
  {98}},\ \bibinfo {pages} {051301} (\bibinfo {year} {2018})}\BibitemShut
  {NoStop}%
\bibitem [{\citenamefont {Litvinova}\ \emph {et~al.}(2014)\citenamefont
  {Litvinova}, \citenamefont {Brown}, \citenamefont {Fang}, \citenamefont
  {Marketin},\ and\ \citenamefont {Zegers}}]{LITVINOVA2014307}%
  \BibitemOpen
  \bibfield  {author} {\bibinfo {author} {\bibfnamefont {E.}~\bibnamefont
  {Litvinova}}, \bibinfo {author} {\bibfnamefont {B.}~\bibnamefont {Brown}},
  \bibinfo {author} {\bibfnamefont {D.-L.}\ \bibnamefont {Fang}}, \bibinfo
  {author} {\bibfnamefont {T.}~\bibnamefont {Marketin}}, \ and\ \bibinfo
  {author} {\bibfnamefont {R.}~\bibnamefont {Zegers}},\ }\href {\doibase
  10.1016/j.physletb.2014.02.001} {\bibfield  {journal} {\bibinfo  {journal}
  {Physics Letters B}\ }\textbf {\bibinfo {volume} {730}},\ \bibinfo {pages}
  {307 } (\bibinfo {year} {2014})}\BibitemShut {NoStop}%
\bibitem [{\citenamefont {Niu}\ \emph {et~al.}(2013{\natexlab{a}})\citenamefont
  {Niu}, \citenamefont {Niu}, \citenamefont {Liang}, \citenamefont {Long},
  \citenamefont {Nik\v{s}i\'{c}}, \citenamefont {Vretenar},\ and\ \citenamefont
  {Meng}}]{NIU2013172}%
  \BibitemOpen
  \bibfield  {author} {\bibinfo {author} {\bibfnamefont {Z.~M.}\ \bibnamefont
  {Niu}}, \bibinfo {author} {\bibfnamefont {Y.~F.}\ \bibnamefont {Niu}},
  \bibinfo {author} {\bibfnamefont {H.~Z.}\ \bibnamefont {Liang}}, \bibinfo
  {author} {\bibfnamefont {W.~H.}\ \bibnamefont {Long}}, \bibinfo {author}
  {\bibfnamefont {T.}~\bibnamefont {Nik\v{s}i\'{c}}}, \bibinfo {author}
  {\bibfnamefont {D.}~\bibnamefont {Vretenar}}, \ and\ \bibinfo {author}
  {\bibfnamefont {J.}~\bibnamefont {Meng}},\ }\href {\doibase
  10.1016/j.physletb.2013.04.048} {\bibfield  {journal} {\bibinfo  {journal}
  {Physics Letters B}\ }\textbf {\bibinfo {volume} {723}},\ \bibinfo {pages}
  {172} (\bibinfo {year} {2013}{\natexlab{a}})}\BibitemShut {NoStop}%
\bibitem [{\citenamefont {Engel}\ \emph {et~al.}(1999)\citenamefont {Engel},
  \citenamefont {Bender}, \citenamefont {Dobaczewski}, \citenamefont
  {Nazarewicz},\ and\ \citenamefont {Surman}}]{PhysRevC.60.014302}%
  \BibitemOpen
  \bibfield  {author} {\bibinfo {author} {\bibfnamefont {J.}~\bibnamefont
  {Engel}}, \bibinfo {author} {\bibfnamefont {M.}~\bibnamefont {Bender}},
  \bibinfo {author} {\bibfnamefont {J.}~\bibnamefont {Dobaczewski}}, \bibinfo
  {author} {\bibfnamefont {W.}~\bibnamefont {Nazarewicz}}, \ and\ \bibinfo
  {author} {\bibfnamefont {R.}~\bibnamefont {Surman}},\ }\href {\doibase
  10.1103/PhysRevC.60.014302} {\bibfield  {journal} {\bibinfo  {journal} {Phys.
  Rev. C}\ }\textbf {\bibinfo {volume} {60}},\ \bibinfo {pages} {014302}
  (\bibinfo {year} {1999})}\BibitemShut {NoStop}%
\bibitem [{\citenamefont {Bender}\ \emph {et~al.}(2002)\citenamefont {Bender},
  \citenamefont {Dobaczewski}, \citenamefont {Engel},\ and\ \citenamefont
  {Nazarewicz}}]{PhysRevC.65.054322}%
  \BibitemOpen
  \bibfield  {author} {\bibinfo {author} {\bibfnamefont {M.}~\bibnamefont
  {Bender}}, \bibinfo {author} {\bibfnamefont {J.}~\bibnamefont {Dobaczewski}},
  \bibinfo {author} {\bibfnamefont {J.}~\bibnamefont {Engel}}, \ and\ \bibinfo
  {author} {\bibfnamefont {W.}~\bibnamefont {Nazarewicz}},\ }\href {\doibase
  10.1103/PhysRevC.65.054322} {\bibfield  {journal} {\bibinfo  {journal} {Phys.
  Rev. C}\ }\textbf {\bibinfo {volume} {65}},\ \bibinfo {pages} {054322}
  (\bibinfo {year} {2002})}\BibitemShut {NoStop}%
\bibitem [{\citenamefont {Fracasso}\ and\ \citenamefont
  {Col\`o}(2005)}]{PhysRevC.72.064310}%
  \BibitemOpen
  \bibfield  {author} {\bibinfo {author} {\bibfnamefont {S.}~\bibnamefont
  {Fracasso}}\ and\ \bibinfo {author} {\bibfnamefont {G.}~\bibnamefont
  {Col\`o}},\ }\href {\doibase 10.1103/PhysRevC.72.064310} {\bibfield
  {journal} {\bibinfo  {journal} {Phys. Rev. C}\ }\textbf {\bibinfo {volume}
  {72}},\ \bibinfo {pages} {064310} (\bibinfo {year} {2005})}\BibitemShut
  {NoStop}%
\bibitem [{\citenamefont {Fracasso}\ and\ \citenamefont
  {Col\`o}(2007)}]{PhysRevC.76.044307}%
  \BibitemOpen
  \bibfield  {author} {\bibinfo {author} {\bibfnamefont {S.}~\bibnamefont
  {Fracasso}}\ and\ \bibinfo {author} {\bibfnamefont {G.}~\bibnamefont
  {Col\`o}},\ }\href {\doibase 10.1103/PhysRevC.76.044307} {\bibfield
  {journal} {\bibinfo  {journal} {Phys. Rev. C}\ }\textbf {\bibinfo {volume}
  {76}},\ \bibinfo {pages} {044307} (\bibinfo {year} {2007})}\BibitemShut
  {NoStop}%
\bibitem [{\citenamefont {Bai}\ \emph {et~al.}(2014)\citenamefont {Bai},
  \citenamefont {Sagawa}, \citenamefont {Col\`o}, \citenamefont {Fujita},
  \citenamefont {Zhang}, \citenamefont {Zhang},\ and\ \citenamefont
  {Xu}}]{PhysRevC.90.054335}%
  \BibitemOpen
  \bibfield  {author} {\bibinfo {author} {\bibfnamefont {C.~L.}\ \bibnamefont
  {Bai}}, \bibinfo {author} {\bibfnamefont {H.}~\bibnamefont {Sagawa}},
  \bibinfo {author} {\bibfnamefont {G.}~\bibnamefont {Col\`o}}, \bibinfo
  {author} {\bibfnamefont {Y.}~\bibnamefont {Fujita}}, \bibinfo {author}
  {\bibfnamefont {H.~Q.}\ \bibnamefont {Zhang}}, \bibinfo {author}
  {\bibfnamefont {X.~Z.}\ \bibnamefont {Zhang}}, \ and\ \bibinfo {author}
  {\bibfnamefont {F.~R.}\ \bibnamefont {Xu}},\ }\href {\doibase
  10.1103/PhysRevC.90.054335} {\bibfield  {journal} {\bibinfo  {journal} {Phys.
  Rev. C}\ }\textbf {\bibinfo {volume} {90}},\ \bibinfo {pages} {054335}
  (\bibinfo {year} {2014})}\BibitemShut {NoStop}%
\bibitem [{\citenamefont {Sarriguren}\ \emph {et~al.}(2018)\citenamefont
  {Sarriguren}, \citenamefont {Algora},\ and\ \citenamefont
  {Kiss}}]{PhysRevC.98.024311}%
  \BibitemOpen
  \bibfield  {author} {\bibinfo {author} {\bibfnamefont {P.}~\bibnamefont
  {Sarriguren}}, \bibinfo {author} {\bibfnamefont {A.}~\bibnamefont {Algora}},
  \ and\ \bibinfo {author} {\bibfnamefont {G.}~\bibnamefont {Kiss}},\ }\href
  {\doibase 10.1103/PhysRevC.98.024311} {\bibfield  {journal} {\bibinfo
  {journal} {Phys. Rev. C}\ }\textbf {\bibinfo {volume} {98}},\ \bibinfo
  {pages} {024311} (\bibinfo {year} {2018})}\BibitemShut {NoStop}%
\bibitem [{\citenamefont {Deloncle}\ \emph {et~al.}(2017)\citenamefont
  {Deloncle}, \citenamefont {P{\'e}ru},\ and\ \citenamefont
  {Martini}}]{Deloncle2017}%
  \BibitemOpen
  \bibfield  {author} {\bibinfo {author} {\bibfnamefont {I.}~\bibnamefont
  {Deloncle}}, \bibinfo {author} {\bibfnamefont {S.}~\bibnamefont {P{\'e}ru}},
  \ and\ \bibinfo {author} {\bibfnamefont {M.}~\bibnamefont {Martini}},\ }\href
  {\doibase 10.1140/epja/i2017-12354-x} {\bibfield  {journal} {\bibinfo
  {journal} {The European Physical Journal A}\ }\textbf {\bibinfo {volume}
  {53}},\ \bibinfo {pages} {170} (\bibinfo {year} {2017})}\BibitemShut
  {NoStop}%
\bibitem [{\citenamefont {Bai}\ \emph {et~al.}(2013)\citenamefont {Bai},
  \citenamefont {Sagawa}, \citenamefont {Sasano}, \citenamefont {Uesaka},
  \citenamefont {Hagino}, \citenamefont {Zhang}, \citenamefont {Zhang},\ and\
  \citenamefont {Xu}}]{BAI2013116}%
  \BibitemOpen
  \bibfield  {author} {\bibinfo {author} {\bibfnamefont {C.~L.}\ \bibnamefont
  {Bai}}, \bibinfo {author} {\bibfnamefont {H.}~\bibnamefont {Sagawa}},
  \bibinfo {author} {\bibfnamefont {M.}~\bibnamefont {Sasano}}, \bibinfo
  {author} {\bibfnamefont {T.}~\bibnamefont {Uesaka}}, \bibinfo {author}
  {\bibfnamefont {K.}~\bibnamefont {Hagino}}, \bibinfo {author} {\bibfnamefont
  {H.~Q.}\ \bibnamefont {Zhang}}, \bibinfo {author} {\bibfnamefont {X.~Z.}\
  \bibnamefont {Zhang}}, \ and\ \bibinfo {author} {\bibfnamefont {F.~R.}\
  \bibnamefont {Xu}},\ }\href {\doibase 10.1016/j.physletb.2012.12.060}
  {\bibfield  {journal} {\bibinfo  {journal} {Physics Letters B}\ }\textbf
  {\bibinfo {volume} {719}},\ \bibinfo {pages} {116} (\bibinfo {year}
  {2013})}\BibitemShut {NoStop}%
\bibitem [{\citenamefont {Martini}\ \emph {et~al.}(2014)\citenamefont
  {Martini}, \citenamefont {P\'eru},\ and\ \citenamefont
  {Goriely}}]{PhysRevC.89.044306}%
  \BibitemOpen
  \bibfield  {author} {\bibinfo {author} {\bibfnamefont {M.}~\bibnamefont
  {Martini}}, \bibinfo {author} {\bibfnamefont {S.}~\bibnamefont {P\'eru}}, \
  and\ \bibinfo {author} {\bibfnamefont {S.}~\bibnamefont {Goriely}},\ }\href
  {\doibase 10.1103/PhysRevC.89.044306} {\bibfield  {journal} {\bibinfo
  {journal} {Phys. Rev. C}\ }\textbf {\bibinfo {volume} {89}},\ \bibinfo
  {pages} {044306} (\bibinfo {year} {2014})}\BibitemShut {NoStop}%
\bibitem [{\citenamefont {Niu}\ \emph {et~al.}(2018)\citenamefont {Niu},
  \citenamefont {Niu}, \citenamefont {Col\`{o}},\ and\ \citenamefont
  {Vigezzi}}]{NIU2018325}%
  \BibitemOpen
  \bibfield  {author} {\bibinfo {author} {\bibfnamefont {Y.~F.}\ \bibnamefont
  {Niu}}, \bibinfo {author} {\bibfnamefont {Z.~M.}\ \bibnamefont {Niu}},
  \bibinfo {author} {\bibfnamefont {G.}~\bibnamefont {Col\`{o}}}, \ and\
  \bibinfo {author} {\bibfnamefont {E.}~\bibnamefont {Vigezzi}},\ }\href
  {\doibase 10.1016/j.physletb.2018.02.061} {\bibfield  {journal} {\bibinfo
  {journal} {Physics Letters B}\ }\textbf {\bibinfo {volume} {780}},\ \bibinfo
  {pages} {325} (\bibinfo {year} {2018})}\BibitemShut {NoStop}%
\bibitem [{\citenamefont {Brown}(2001)}]{BROWN2001517}%
  \BibitemOpen
  \bibfield  {author} {\bibinfo {author} {\bibfnamefont {B.~A.}\ \bibnamefont
  {Brown}},\ }\href {\doibase 10.1016/S0146-6410(01)00159-4} {\bibfield
  {journal} {\bibinfo  {journal} {Progress in Particle and Nuclear Physics}\
  }\textbf {\bibinfo {volume} {47}},\ \bibinfo {pages} {517 } (\bibinfo {year}
  {2001})}\BibitemShut {NoStop}%
\bibitem [{\citenamefont {Caurier}\ \emph {et~al.}(2005)\citenamefont
  {Caurier}, \citenamefont {Mart\'{\i}nez-Pinedo}, \citenamefont {Nowacki},
  \citenamefont {Poves},\ and\ \citenamefont {Zuker}}]{RevModPhys.77.427}%
  \BibitemOpen
  \bibfield  {author} {\bibinfo {author} {\bibfnamefont {E.}~\bibnamefont
  {Caurier}}, \bibinfo {author} {\bibfnamefont {G.}~\bibnamefont
  {Mart\'{\i}nez-Pinedo}}, \bibinfo {author} {\bibfnamefont {F.}~\bibnamefont
  {Nowacki}}, \bibinfo {author} {\bibfnamefont {A.}~\bibnamefont {Poves}}, \
  and\ \bibinfo {author} {\bibfnamefont {A.~P.}\ \bibnamefont {Zuker}},\ }\href
  {\doibase 10.1103/RevModPhys.77.427} {\bibfield  {journal} {\bibinfo
  {journal} {Rev. Mod. Phys.}\ }\textbf {\bibinfo {volume} {77}},\ \bibinfo
  {pages} {427} (\bibinfo {year} {2005})}\BibitemShut {NoStop}%
\bibitem [{\citenamefont {Paar}\ \emph {et~al.}(2007)\citenamefont {Paar},
  \citenamefont {Vretenar}, \citenamefont {Khan},\ and\ \citenamefont
  {Col{\`{o}}}}]{Paar_2007}%
  \BibitemOpen
  \bibfield  {author} {\bibinfo {author} {\bibfnamefont {N.}~\bibnamefont
  {Paar}}, \bibinfo {author} {\bibfnamefont {D.}~\bibnamefont {Vretenar}},
  \bibinfo {author} {\bibfnamefont {E.}~\bibnamefont {Khan}}, \ and\ \bibinfo
  {author} {\bibfnamefont {G.}~\bibnamefont {Col{\`{o}}}},\ }\href {\doibase
  10.1088/0034-4885/70/5/r02} {\bibfield  {journal} {\bibinfo  {journal}
  {Reports on Progress in Physics}\ }\textbf {\bibinfo {volume} {70}},\
  \bibinfo {pages} {691} (\bibinfo {year} {2007})}\BibitemShut {NoStop}%
\bibitem [{\citenamefont {Bender}\ \emph {et~al.}(2000)\citenamefont {Bender},
  \citenamefont {Rutz}, \citenamefont {Reinhard},\ and\ \citenamefont
  {Maruhn}}]{Bender2000}%
  \BibitemOpen
  \bibfield  {author} {\bibinfo {author} {\bibfnamefont {M.}~\bibnamefont
  {Bender}}, \bibinfo {author} {\bibfnamefont {K.}~\bibnamefont {Rutz}},
  \bibinfo {author} {\bibfnamefont {P.~G.}\ \bibnamefont {Reinhard}}, \ and\
  \bibinfo {author} {\bibfnamefont {J.~A.}\ \bibnamefont {Maruhn}},\ }\href
  {\doibase 10.1007/s10050-000-4504-z} {\bibfield  {journal} {\bibinfo
  {journal} {The European Physical Journal A}\ }\textbf {\bibinfo {volume}
  {8}},\ \bibinfo {pages} {59} (\bibinfo {year} {2000})}\BibitemShut {NoStop}%
\bibitem [{\citenamefont {Changizi}\ \emph {et~al.}(2015)\citenamefont
  {Changizi}, \citenamefont {Qi},\ and\ \citenamefont
  {Wyss}}]{CHANGIZI2015210}%
  \BibitemOpen
  \bibfield  {author} {\bibinfo {author} {\bibfnamefont {S.}~\bibnamefont
  {Changizi}}, \bibinfo {author} {\bibfnamefont {C.}~\bibnamefont {Qi}}, \ and\
  \bibinfo {author} {\bibfnamefont {R.}~\bibnamefont {Wyss}},\ }\href {\doibase
  10.1016/j.nuclphysa.2015.04.010} {\bibfield  {journal} {\bibinfo  {journal}
  {Nuclear Physics A}\ }\textbf {\bibinfo {volume} {940}},\ \bibinfo {pages}
  {210 } (\bibinfo {year} {2015})}\BibitemShut {NoStop}%
\bibitem [{\citenamefont {Goodman}(1999)}]{PhysRevC.60.014311}%
  \BibitemOpen
  \bibfield  {author} {\bibinfo {author} {\bibfnamefont {A.~L.}\ \bibnamefont
  {Goodman}},\ }\href {\doibase 10.1103/PhysRevC.60.014311} {\bibfield
  {journal} {\bibinfo  {journal} {Phys. Rev. C}\ }\textbf {\bibinfo {volume}
  {60}},\ \bibinfo {pages} {014311} (\bibinfo {year} {1999})}\BibitemShut
  {NoStop}%
\bibitem [{\citenamefont {Langanke}\ \emph {et~al.}(1997)\citenamefont
  {Langanke}, \citenamefont {Dean}, \citenamefont {Koonin},\ and\ \citenamefont
  {Radha}}]{LANGANKE1997253}%
  \BibitemOpen
  \bibfield  {author} {\bibinfo {author} {\bibfnamefont {K.}~\bibnamefont
  {Langanke}}, \bibinfo {author} {\bibfnamefont {D.}~\bibnamefont {Dean}},
  \bibinfo {author} {\bibfnamefont {S.}~\bibnamefont {Koonin}}, \ and\ \bibinfo
  {author} {\bibfnamefont {P.}~\bibnamefont {Radha}},\ }\href {\doibase
  10.1016/S0375-9474(96)00442-3} {\bibfield  {journal} {\bibinfo  {journal}
  {Nuclear Physics A}\ }\textbf {\bibinfo {volume} {613}},\ \bibinfo {pages}
  {253 } (\bibinfo {year} {1997})}\BibitemShut {NoStop}%
\bibitem [{\citenamefont {Mart\'{\i}nez-Pinedo}\ \emph
  {et~al.}(1999)\citenamefont {Mart\'{\i}nez-Pinedo}, \citenamefont
  {Langanke},\ and\ \citenamefont {Vogel}}]{MARTINEZPINEDO1999379}%
  \BibitemOpen
  \bibfield  {author} {\bibinfo {author} {\bibfnamefont {G.}~\bibnamefont
  {Mart\'{\i}nez-Pinedo}}, \bibinfo {author} {\bibfnamefont {K.}~\bibnamefont
  {Langanke}}, \ and\ \bibinfo {author} {\bibfnamefont {P.}~\bibnamefont
  {Vogel}},\ }\href {\doibase 10.1016/S0375-9474(99)00141-4} {\bibfield
  {journal} {\bibinfo  {journal} {Nuclear Physics A}\ }\textbf {\bibinfo
  {volume} {651}},\ \bibinfo {pages} {379 } (\bibinfo {year}
  {1999})}\BibitemShut {NoStop}%
\bibitem [{\citenamefont {Poves}\ and\ \citenamefont
  {Mart\'{\i}nez-Pinedo}(1998)}]{POVES1998203}%
  \BibitemOpen
  \bibfield  {author} {\bibinfo {author} {\bibfnamefont {A.}~\bibnamefont
  {Poves}}\ and\ \bibinfo {author} {\bibfnamefont {G.}~\bibnamefont
  {Mart\'{\i}nez-Pinedo}},\ }\href {\doibase 10.1016/S0370-2693(98)00538-3}
  {\bibfield  {journal} {\bibinfo  {journal} {Physics Letters B}\ }\textbf
  {\bibinfo {volume} {430}},\ \bibinfo {pages} {203 } (\bibinfo {year}
  {1998})}\BibitemShut {NoStop}%
\bibitem [{\citenamefont {Warner}\ \emph {et~al.}(2006)\citenamefont {Warner},
  \citenamefont {Bentley},\ and\ \citenamefont {Van~Isacker}}]{warner2006}%
  \BibitemOpen
  \bibfield  {author} {\bibinfo {author} {\bibfnamefont {D.~D.}\ \bibnamefont
  {Warner}}, \bibinfo {author} {\bibfnamefont {M.~A.}\ \bibnamefont {Bentley}},
  \ and\ \bibinfo {author} {\bibfnamefont {P.}~\bibnamefont {Van~Isacker}},\
  }\href {\doibase 10.1038/nphys291} {\bibfield  {journal} {\bibinfo  {journal}
  {Nature Physics}\ }\textbf {\bibinfo {volume} {2}},\ \bibinfo {pages} {311}
  (\bibinfo {year} {2006})}\BibitemShut {NoStop}%
\bibitem [{\citenamefont {Sambataro}\ \emph {et~al.}(2015)\citenamefont
  {Sambataro}, \citenamefont {Sandulescu},\ and\ \citenamefont
  {Johnson}}]{SAMBATARO2015137}%
  \BibitemOpen
  \bibfield  {author} {\bibinfo {author} {\bibfnamefont {M.}~\bibnamefont
  {Sambataro}}, \bibinfo {author} {\bibfnamefont {N.}~\bibnamefont
  {Sandulescu}}, \ and\ \bibinfo {author} {\bibfnamefont {C.}~\bibnamefont
  {Johnson}},\ }\href {\doibase 10.1016/j.physletb.2014.11.036} {\bibfield
  {journal} {\bibinfo  {journal} {Physics Letters B}\ }\textbf {\bibinfo
  {volume} {740}},\ \bibinfo {pages} {137} (\bibinfo {year}
  {2015})}\BibitemShut {NoStop}%
\bibitem [{\citenamefont {Negrea}\ \emph {et~al.}(2018)\citenamefont {Negrea},
  \citenamefont {Buganu}, \citenamefont {Gambacurta},\ and\ \citenamefont
  {Sandulescu}}]{PhysRevC.98.064319}%
  \BibitemOpen
  \bibfield  {author} {\bibinfo {author} {\bibfnamefont {D.}~\bibnamefont
  {Negrea}}, \bibinfo {author} {\bibfnamefont {P.}~\bibnamefont {Buganu}},
  \bibinfo {author} {\bibfnamefont {D.}~\bibnamefont {Gambacurta}}, \ and\
  \bibinfo {author} {\bibfnamefont {N.}~\bibnamefont {Sandulescu}},\ }\href
  {\doibase 10.1103/PhysRevC.98.064319} {\bibfield  {journal} {\bibinfo
  {journal} {Phys. Rev. C}\ }\textbf {\bibinfo {volume} {98}},\ \bibinfo
  {pages} {064319} (\bibinfo {year} {2018})}\BibitemShut {NoStop}%
\bibitem [{\citenamefont {Frauendorf}\ and\ \citenamefont
  {Macchiavelli}(2014)}]{FRAUENDORF201424}%
  \BibitemOpen
  \bibfield  {author} {\bibinfo {author} {\bibfnamefont {S.}~\bibnamefont
  {Frauendorf}}\ and\ \bibinfo {author} {\bibfnamefont {A.}~\bibnamefont
  {Macchiavelli}},\ }\href {\doibase 10.1016/j.ppnp.2014.07.001} {\bibfield
  {journal} {\bibinfo  {journal} {Progress in Particle and Nuclear Physics}\
  }\textbf {\bibinfo {volume} {78}},\ \bibinfo {pages} {24 } (\bibinfo {year}
  {2014})}\BibitemShut {NoStop}%
\bibitem [{\citenamefont {Sagawa}\ \emph {et~al.}(2016)\citenamefont {Sagawa},
  \citenamefont {Bai},\ and\ \citenamefont {Col{\`{o}}}}]{Sagawa_2016}%
  \BibitemOpen
  \bibfield  {author} {\bibinfo {author} {\bibfnamefont {H.}~\bibnamefont
  {Sagawa}}, \bibinfo {author} {\bibfnamefont {C.~L.}\ \bibnamefont {Bai}}, \
  and\ \bibinfo {author} {\bibfnamefont {G.}~\bibnamefont {Col{\`{o}}}},\
  }\href {\doibase 10.1088/0031-8949/91/8/083011} {\bibfield  {journal}
  {\bibinfo  {journal} {Physica Scripta}\ }\textbf {\bibinfo {volume} {91}},\
  \bibinfo {pages} {083011} (\bibinfo {year} {2016})}\BibitemShut {NoStop}%
\bibitem [{\citenamefont {Goodman}(1981)}]{GOODMAN198130}%
  \BibitemOpen
  \bibfield  {author} {\bibinfo {author} {\bibfnamefont {A.~L.}\ \bibnamefont
  {Goodman}},\ }\href {\doibase 10.1016/0375-9474(81)90557-1} {\bibfield
  {journal} {\bibinfo  {journal} {Nuclear Physics A}\ }\textbf {\bibinfo
  {volume} {352}},\ \bibinfo {pages} {30 } (\bibinfo {year}
  {1981})}\BibitemShut {NoStop}%
\bibitem [{\citenamefont {Khan}\ \emph {et~al.}(2007)\citenamefont {Khan},
  \citenamefont {Giai},\ and\ \citenamefont {Sandulescu}}]{KHAN200794}%
  \BibitemOpen
  \bibfield  {author} {\bibinfo {author} {\bibfnamefont {E.}~\bibnamefont
  {Khan}}, \bibinfo {author} {\bibfnamefont {N.~V.}\ \bibnamefont {Giai}}, \
  and\ \bibinfo {author} {\bibfnamefont {N.}~\bibnamefont {Sandulescu}},\
  }\href {\doibase 10.1016/j.nuclphysa.2007.03.005} {\bibfield  {journal}
  {\bibinfo  {journal} {Nuclear Physics A}\ }\textbf {\bibinfo {volume}
  {789}},\ \bibinfo {pages} {94 } (\bibinfo {year} {2007})}\BibitemShut
  {NoStop}%
\bibitem [{\citenamefont {Niu}\ \emph {et~al.}(2013{\natexlab{b}})\citenamefont
  {Niu}, \citenamefont {Niu}, \citenamefont {Paar}, \citenamefont {Vretenar},
  \citenamefont {Wang}, \citenamefont {Bai},\ and\ \citenamefont
  {Meng}}]{PhysRevC.88.034308}%
  \BibitemOpen
  \bibfield  {author} {\bibinfo {author} {\bibfnamefont {Y.~F.}\ \bibnamefont
  {Niu}}, \bibinfo {author} {\bibfnamefont {Z.~M.}\ \bibnamefont {Niu}},
  \bibinfo {author} {\bibfnamefont {N.}~\bibnamefont {Paar}}, \bibinfo {author}
  {\bibfnamefont {D.}~\bibnamefont {Vretenar}}, \bibinfo {author}
  {\bibfnamefont {G.~H.}\ \bibnamefont {Wang}}, \bibinfo {author}
  {\bibfnamefont {J.~S.}\ \bibnamefont {Bai}}, \ and\ \bibinfo {author}
  {\bibfnamefont {J.}~\bibnamefont {Meng}},\ }\href {\doibase
  10.1103/PhysRevC.88.034308} {\bibfield  {journal} {\bibinfo  {journal} {Phys.
  Rev. C}\ }\textbf {\bibinfo {volume} {88}},\ \bibinfo {pages} {034308}
  (\bibinfo {year} {2013}{\natexlab{b}})}\BibitemShut {NoStop}%
\bibitem [{\citenamefont {{Y\"uksel}}\ \emph {et~al.}(2014)\citenamefont
  {{Y\"uksel}}, \citenamefont {{E. Khan}}, \citenamefont {{K. Bozkurt}},\ and\
  \citenamefont {{G. Col\`o}}}]{refId0}%
  \BibitemOpen
  \bibfield  {author} {\bibinfo {author} {\bibfnamefont {E.}~\bibnamefont
  {{Y\"uksel}}}, \bibinfo {author} {\bibnamefont {{E. Khan}}}, \bibinfo
  {author} {\bibnamefont {{K. Bozkurt}}}, \ and\ \bibinfo {author}
  {\bibnamefont {{G. Col\`o}}},\ }\href {\doibase 10.1140/epja/i2014-14160-4}
  {\bibfield  {journal} {\bibinfo  {journal} {Eur. Phys. J. A}\ }\textbf
  {\bibinfo {volume} {50}},\ \bibinfo {pages} {160} (\bibinfo {year}
  {2014})}\BibitemShut {NoStop}%
\bibitem [{\citenamefont {Li}\ \emph {et~al.}(2015)\citenamefont {Li},
  \citenamefont {Margueron}, \citenamefont {Long},\ and\ \citenamefont
  {Van~Giai}}]{PhysRevC.92.014302}%
  \BibitemOpen
  \bibfield  {author} {\bibinfo {author} {\bibfnamefont {J.~J.}\ \bibnamefont
  {Li}}, \bibinfo {author} {\bibfnamefont {J.}~\bibnamefont {Margueron}},
  \bibinfo {author} {\bibfnamefont {W.~H.}\ \bibnamefont {Long}}, \ and\
  \bibinfo {author} {\bibfnamefont {N.}~\bibnamefont {Van~Giai}},\ }\href
  {\doibase 10.1103/PhysRevC.92.014302} {\bibfield  {journal} {\bibinfo
  {journal} {Phys. Rev. C}\ }\textbf {\bibinfo {volume} {92}},\ \bibinfo
  {pages} {014302} (\bibinfo {year} {2015})}\BibitemShut {NoStop}%
\bibitem [{\citenamefont {Belabbas}\ \emph {et~al.}(2017)\citenamefont
  {Belabbas}, \citenamefont {Li},\ and\ \citenamefont
  {Margueron}}]{PhysRevC.96.024304}%
  \BibitemOpen
  \bibfield  {author} {\bibinfo {author} {\bibfnamefont {M.}~\bibnamefont
  {Belabbas}}, \bibinfo {author} {\bibfnamefont {J.~J.}\ \bibnamefont {Li}}, \
  and\ \bibinfo {author} {\bibfnamefont {J.}~\bibnamefont {Margueron}},\ }\href
  {\doibase 10.1103/PhysRevC.96.024304} {\bibfield  {journal} {\bibinfo
  {journal} {Phys. Rev. C}\ }\textbf {\bibinfo {volume} {96}},\ \bibinfo
  {pages} {024304} (\bibinfo {year} {2017})}\BibitemShut {NoStop}%
\bibitem [{\citenamefont {Zhang}\ and\ \citenamefont
  {Niu}(2018)}]{PhysRevC.97.054302}%
  \BibitemOpen
  \bibfield  {author} {\bibinfo {author} {\bibfnamefont {W.}~\bibnamefont
  {Zhang}}\ and\ \bibinfo {author} {\bibfnamefont {Y.~F.}\ \bibnamefont
  {Niu}},\ }\href {\doibase 10.1103/PhysRevC.97.054302} {\bibfield  {journal}
  {\bibinfo  {journal} {Phys. Rev. C}\ }\textbf {\bibinfo {volume} {97}},\
  \bibinfo {pages} {054302} (\bibinfo {year} {2018})}\BibitemShut {NoStop}%
\bibitem [{\citenamefont {Sommermann}(1983)}]{SOMMERMANN1983163}%
  \BibitemOpen
  \bibfield  {author} {\bibinfo {author} {\bibfnamefont {H.}~\bibnamefont
  {Sommermann}},\ }\href {\doibase 10.1016/0003-4916(83)90318-4} {\bibfield
  {journal} {\bibinfo  {journal} {Annals of Physics}\ }\textbf {\bibinfo
  {volume} {151}},\ \bibinfo {pages} {163 } (\bibinfo {year}
  {1983})}\BibitemShut {NoStop}%
\bibitem [{\citenamefont {Khan}\ \emph {et~al.}(2004)\citenamefont {Khan},
  \citenamefont {Giai},\ and\ \citenamefont {Grasso}}]{KHAN2004311}%
  \BibitemOpen
  \bibfield  {author} {\bibinfo {author} {\bibfnamefont {E.}~\bibnamefont
  {Khan}}, \bibinfo {author} {\bibfnamefont {N.~V.}\ \bibnamefont {Giai}}, \
  and\ \bibinfo {author} {\bibfnamefont {M.}~\bibnamefont {Grasso}},\ }\href
  {\doibase 10.1016/j.nuclphysa.2003.11.042} {\bibfield  {journal} {\bibinfo
  {journal} {Nuclear Physics A}\ }\textbf {\bibinfo {volume} {731}},\ \bibinfo
  {pages} {311 } (\bibinfo {year} {2004})}\BibitemShut {NoStop}%
\bibitem [{\citenamefont {Y\"uksel}\ \emph {et~al.}(2017)\citenamefont
  {Y\"uksel}, \citenamefont {Col\`o}, \citenamefont {Khan}, \citenamefont
  {Niu},\ and\ \citenamefont {Bozkurt}}]{PhysRevC.96.024303}%
  \BibitemOpen
  \bibfield  {author} {\bibinfo {author} {\bibfnamefont {E.}~\bibnamefont
  {Y\"uksel}}, \bibinfo {author} {\bibfnamefont {G.}~\bibnamefont {Col\`o}},
  \bibinfo {author} {\bibfnamefont {E.}~\bibnamefont {Khan}}, \bibinfo {author}
  {\bibfnamefont {Y.~F.}\ \bibnamefont {Niu}}, \ and\ \bibinfo {author}
  {\bibfnamefont {K.}~\bibnamefont {Bozkurt}},\ }\href {\doibase
  10.1103/PhysRevC.96.024303} {\bibfield  {journal} {\bibinfo  {journal} {Phys.
  Rev. C}\ }\textbf {\bibinfo {volume} {96}},\ \bibinfo {pages} {024303}
  (\bibinfo {year} {2017})}\BibitemShut {NoStop}%
\bibitem [{\citenamefont {Langanke}\ \emph {et~al.}(2001)\citenamefont
  {Langanke}, \citenamefont {Kolbe},\ and\ \citenamefont
  {Dean}}]{PhysRevC.63.032801}%
  \BibitemOpen
  \bibfield  {author} {\bibinfo {author} {\bibfnamefont {K.}~\bibnamefont
  {Langanke}}, \bibinfo {author} {\bibfnamefont {E.}~\bibnamefont {Kolbe}}, \
  and\ \bibinfo {author} {\bibfnamefont {D.~J.}\ \bibnamefont {Dean}},\ }\href
  {\doibase 10.1103/PhysRevC.63.032801} {\bibfield  {journal} {\bibinfo
  {journal} {Phys. Rev. C}\ }\textbf {\bibinfo {volume} {63}},\ \bibinfo
  {pages} {032801(R)} (\bibinfo {year} {2001})}\BibitemShut {NoStop}%
\bibitem [{\citenamefont {Dzhioev}\ \emph {et~al.}(2010)\citenamefont
  {Dzhioev}, \citenamefont {Vdovin}, \citenamefont {Ponomarev}, \citenamefont
  {Wambach}, \citenamefont {Langanke},\ and\ \citenamefont
  {Mart\'{\i}nez-Pinedo}}]{PhysRevC.81.015804}%
  \BibitemOpen
  \bibfield  {author} {\bibinfo {author} {\bibfnamefont {A.~A.}\ \bibnamefont
  {Dzhioev}}, \bibinfo {author} {\bibfnamefont {A.~I.}\ \bibnamefont {Vdovin}},
  \bibinfo {author} {\bibfnamefont {V.~Y.}\ \bibnamefont {Ponomarev}}, \bibinfo
  {author} {\bibfnamefont {J.}~\bibnamefont {Wambach}}, \bibinfo {author}
  {\bibfnamefont {K.}~\bibnamefont {Langanke}}, \ and\ \bibinfo {author}
  {\bibfnamefont {G.}~\bibnamefont {Mart\'{\i}nez-Pinedo}},\ }\href {\doibase
  10.1103/PhysRevC.81.015804} {\bibfield  {journal} {\bibinfo  {journal} {Phys.
  Rev. C}\ }\textbf {\bibinfo {volume} {81}},\ \bibinfo {pages} {015804}
  (\bibinfo {year} {2010})}\BibitemShut {NoStop}%
\bibitem [{\citenamefont {Martini}\ \emph {et~al.}(2009)\citenamefont
  {Martini}, \citenamefont {P\'eru},\ and\ \citenamefont
  {Goriely}}]{NIU2009315}%
  \BibitemOpen
  \bibfield  {author} {\bibinfo {author} {\bibfnamefont {M.}~\bibnamefont
  {Martini}}, \bibinfo {author} {\bibfnamefont {S.}~\bibnamefont {P\'eru}}, \
  and\ \bibinfo {author} {\bibfnamefont {S.}~\bibnamefont {Goriely}},\ }\href
  {\doibase 10.1016/j.physletb.2009.10.046} {\bibfield  {journal} {\bibinfo
  {journal} {Physics Letters B}\ }\textbf {\bibinfo {volume} {681}},\ \bibinfo
  {pages} {315} (\bibinfo {year} {2009})}\BibitemShut {NoStop}%
\bibitem [{\citenamefont {Civitarese}\ and\ \citenamefont
  {Reboiro}(2001)}]{PhysRevC.63.034323}%
  \BibitemOpen
  \bibfield  {author} {\bibinfo {author} {\bibfnamefont {O.}~\bibnamefont
  {Civitarese}}\ and\ \bibinfo {author} {\bibfnamefont {M.}~\bibnamefont
  {Reboiro}},\ }\href {\doibase 10.1103/PhysRevC.63.034323} {\bibfield
  {journal} {\bibinfo  {journal} {Phys. Rev. C}\ }\textbf {\bibinfo {volume}
  {63}},\ \bibinfo {pages} {034323} (\bibinfo {year} {2001})}\BibitemShut
  {NoStop}%
\bibitem [{\citenamefont {Niu}\ \emph {et~al.}(2011)\citenamefont {Niu},
  \citenamefont {Paar}, \citenamefont {Vretenar},\ and\ \citenamefont
  {Meng}}]{PhysRevC.83.045807}%
  \BibitemOpen
  \bibfield  {author} {\bibinfo {author} {\bibfnamefont {Y.~F.}\ \bibnamefont
  {Niu}}, \bibinfo {author} {\bibfnamefont {N.}~\bibnamefont {Paar}}, \bibinfo
  {author} {\bibfnamefont {D.}~\bibnamefont {Vretenar}}, \ and\ \bibinfo
  {author} {\bibfnamefont {J.}~\bibnamefont {Meng}},\ }\href {\doibase
  10.1103/PhysRevC.83.045807} {\bibfield  {journal} {\bibinfo  {journal} {Phys.
  Rev. C}\ }\textbf {\bibinfo {volume} {83}},\ \bibinfo {pages} {045807}
  (\bibinfo {year} {2011})}\BibitemShut {NoStop}%
\bibitem [{\citenamefont {Paar}\ \emph {et~al.}(2009)\citenamefont {Paar},
  \citenamefont {Col\`o}, \citenamefont {Khan},\ and\ \citenamefont
  {Vretenar}}]{PhysRevC.80.055801}%
  \BibitemOpen
  \bibfield  {author} {\bibinfo {author} {\bibfnamefont {N.}~\bibnamefont
  {Paar}}, \bibinfo {author} {\bibfnamefont {G.}~\bibnamefont {Col\`o}},
  \bibinfo {author} {\bibfnamefont {E.}~\bibnamefont {Khan}}, \ and\ \bibinfo
  {author} {\bibfnamefont {D.}~\bibnamefont {Vretenar}},\ }\href {\doibase
  10.1103/PhysRevC.80.055801} {\bibfield  {journal} {\bibinfo  {journal} {Phys.
  Rev. C}\ }\textbf {\bibinfo {volume} {80}},\ \bibinfo {pages} {055801}
  (\bibinfo {year} {2009})}\BibitemShut {NoStop}%
\bibitem [{\citenamefont {Litvinova}\ and\ \citenamefont
  {Wibowo}(2018)}]{PhysRevLett.121.082501}%
  \BibitemOpen
  \bibfield  {author} {\bibinfo {author} {\bibfnamefont {E.}~\bibnamefont
  {Litvinova}}\ and\ \bibinfo {author} {\bibfnamefont {H.}~\bibnamefont
  {Wibowo}},\ }\href {\doibase 10.1103/PhysRevLett.121.082501} {\bibfield
  {journal} {\bibinfo  {journal} {Phys. Rev. Lett.}\ }\textbf {\bibinfo
  {volume} {121}},\ \bibinfo {pages} {082501} (\bibinfo {year}
  {2018})}\BibitemShut {NoStop}%
\bibitem [{\citenamefont {Dzhioev}\ \emph {et~al.}(2016)\citenamefont
  {Dzhioev}, \citenamefont {Vdovin}, \citenamefont {Mart\'{\i}nez-Pinedo},
  \citenamefont {Wambach},\ and\ \citenamefont
  {Stoyanov}}]{PhysRevC.94.015805}%
  \BibitemOpen
  \bibfield  {author} {\bibinfo {author} {\bibfnamefont {A.~A.}\ \bibnamefont
  {Dzhioev}}, \bibinfo {author} {\bibfnamefont {A.~I.}\ \bibnamefont {Vdovin}},
  \bibinfo {author} {\bibfnamefont {G.}~\bibnamefont {Mart\'{\i}nez-Pinedo}},
  \bibinfo {author} {\bibfnamefont {J.}~\bibnamefont {Wambach}}, \ and\
  \bibinfo {author} {\bibfnamefont {C.}~\bibnamefont {Stoyanov}},\ }\href
  {\doibase 10.1103/PhysRevC.94.015805} {\bibfield  {journal} {\bibinfo
  {journal} {Phys. Rev. C}\ }\textbf {\bibinfo {volume} {94}},\ \bibinfo
  {pages} {015805} (\bibinfo {year} {2016})}\BibitemShut {NoStop}%
\bibitem [{\citenamefont {Litvinova}\ \emph {et~al.}(2018)\citenamefont
  {Litvinova}, \citenamefont {Robin},\ and\ \citenamefont {Wibowo}}]{lit18}%
  \BibitemOpen
  \bibfield  {author} {\bibinfo {author} {\bibfnamefont {E.}~\bibnamefont
  {Litvinova}}, \bibinfo {author} {\bibfnamefont {C.}~\bibnamefont {Robin}}, \
  and\ \bibinfo {author} {\bibfnamefont {H.}~\bibnamefont {Wibowo}},\ }\href
  {https://arxiv.org/abs/1808.07223} {\bibfield  {journal} {\bibinfo  {journal}
  {arXiv:1808.07223 [nucl-th]}\ } (\bibinfo {year} {2018})}\BibitemShut
  {NoStop}%
\bibitem [{\citenamefont {Bartel}\ \emph {et~al.}(1982)\citenamefont {Bartel},
  \citenamefont {Quentin}, \citenamefont {Brack}, \citenamefont {Guet},\ and\
  \citenamefont {H\r{a}kansson}}]{SKM}%
  \BibitemOpen
  \bibfield  {author} {\bibinfo {author} {\bibfnamefont {J.}~\bibnamefont
  {Bartel}}, \bibinfo {author} {\bibfnamefont {P.}~\bibnamefont {Quentin}},
  \bibinfo {author} {\bibfnamefont {M.}~\bibnamefont {Brack}}, \bibinfo
  {author} {\bibfnamefont {C.}~\bibnamefont {Guet}}, \ and\ \bibinfo {author}
  {\bibfnamefont {H.-B.}\ \bibnamefont {H\r{a}kansson}},\ }\href {\doibase
  10.1016/0375-9474(82)90403-1} {\bibfield  {journal} {\bibinfo  {journal}
  {Nuclear Physics A}\ }\textbf {\bibinfo {volume} {386}},\ \bibinfo {pages}
  {79 } (\bibinfo {year} {1982})}\BibitemShut {NoStop}%
\bibitem [{\citenamefont {Lalazissis}\ \emph {et~al.}(2005)\citenamefont
  {Lalazissis}, \citenamefont {Nik\ifmmode \check{s}\else
  \v{s}\fi{}i\ifmmode~\acute{c}\else \'{c}\fi{}}, \citenamefont {Vretenar},\
  and\ \citenamefont {Ring}}]{DD-ME2}%
  \BibitemOpen
  \bibfield  {author} {\bibinfo {author} {\bibfnamefont {G.~A.}\ \bibnamefont
  {Lalazissis}}, \bibinfo {author} {\bibfnamefont {T.}~\bibnamefont
  {Nik\ifmmode \check{s}\else \v{s}\fi{}i\ifmmode~\acute{c}\else \'{c}\fi{}}},
  \bibinfo {author} {\bibfnamefont {D.}~\bibnamefont {Vretenar}}, \ and\
  \bibinfo {author} {\bibfnamefont {P.}~\bibnamefont {Ring}},\ }\href {\doibase
  10.1103/PhysRevC.71.024312} {\bibfield  {journal} {\bibinfo  {journal} {Phys.
  Rev. C}\ }\textbf {\bibinfo {volume} {71}},\ \bibinfo {pages} {024312}
  (\bibinfo {year} {2005})}\BibitemShut {NoStop}%
\bibitem [{\citenamefont {Niu}\ \emph {et~al.}(2016)\citenamefont {Niu},
  \citenamefont {Col\`o}, \citenamefont {Vigezzi}, \citenamefont {Bai},\ and\
  \citenamefont {Sagawa}}]{PhysRevC.94.064328}%
  \BibitemOpen
  \bibfield  {author} {\bibinfo {author} {\bibfnamefont {Y.~F.}\ \bibnamefont
  {Niu}}, \bibinfo {author} {\bibfnamefont {G.}~\bibnamefont {Col\`o}},
  \bibinfo {author} {\bibfnamefont {E.}~\bibnamefont {Vigezzi}}, \bibinfo
  {author} {\bibfnamefont {C.~L.}\ \bibnamefont {Bai}}, \ and\ \bibinfo
  {author} {\bibfnamefont {H.}~\bibnamefont {Sagawa}},\ }\href {\doibase
  10.1103/PhysRevC.94.064328} {\bibfield  {journal} {\bibinfo  {journal} {Phys.
  Rev. C}\ }\textbf {\bibinfo {volume} {94}},\ \bibinfo {pages} {064328}
  (\bibinfo {year} {2016})}\BibitemShut {NoStop}%
\bibitem [{\citenamefont {Ikeda}\ \emph {et~al.}(1963)\citenamefont {Ikeda},
  \citenamefont {Fujii},\ and\ \citenamefont {Fujita}}]{IKEDA1963271}%
  \BibitemOpen
  \bibfield  {author} {\bibinfo {author} {\bibfnamefont {K.}~\bibnamefont
  {Ikeda}}, \bibinfo {author} {\bibfnamefont {S.}~\bibnamefont {Fujii}}, \ and\
  \bibinfo {author} {\bibfnamefont {J.}~\bibnamefont {Fujita}},\ }\href
  {\doibase 10.1016/0031-9163(63)90255-5} {\bibfield  {journal} {\bibinfo
  {journal} {Physics Letters}\ }\textbf {\bibinfo {volume} {3}},\ \bibinfo
  {pages} {271 } (\bibinfo {year} {1963})}\BibitemShut {NoStop}%
\bibitem [{\citenamefont {Fujita}\ \emph {et~al.}(2014)\citenamefont {Fujita},
  \citenamefont {Fujita}, \citenamefont {Adachi}, \citenamefont {Bai},
  \citenamefont {Algora}, \citenamefont {Berg}, \citenamefont {von Brentano},
  \citenamefont {Col\`o}, \citenamefont {Csatl\'os}, \citenamefont {Deaven},
  \citenamefont {Estevez-Aguado}, \citenamefont {Fransen}, \citenamefont
  {De~Frenne}, \citenamefont {Fujita}, \citenamefont
  {Ganio\ifmmode~\breve{g}\else \u{g}\fi{}lu}, \citenamefont {Guess},
  \citenamefont {Guly\'as}, \citenamefont {Hatanaka}, \citenamefont {Hirota},
  \citenamefont {Honma}, \citenamefont {Ishikawa}, \citenamefont {Jacobs},
  \citenamefont {Krasznahorkay}, \citenamefont {Matsubara}, \citenamefont
  {Matsuyanagi}, \citenamefont {Meharchand}, \citenamefont {Molina},
  \citenamefont {Muto}, \citenamefont {Nakanishi}, \citenamefont {Negret},
  \citenamefont {Okamura}, \citenamefont {Ong}, \citenamefont {Otsuka},
  \citenamefont {Pietralla}, \citenamefont {Perdikakis}, \citenamefont
  {Popescu}, \citenamefont {Rubio}, \citenamefont {Sagawa}, \citenamefont
  {Sarriguren}, \citenamefont {Scholl}, \citenamefont {Shimbara}, \citenamefont
  {Shimizu}, \citenamefont {Susoy}, \citenamefont {Suzuki}, \citenamefont
  {Tameshige}, \citenamefont {Tamii}, \citenamefont {Thies}, \citenamefont
  {Uchida}, \citenamefont {Wakasa}, \citenamefont {Yosoi}, \citenamefont
  {Zegers}, \citenamefont {Zell},\ and\ \citenamefont
  {Zenihiro}}]{PhysRevLett.112.112502}%
  \BibitemOpen
  \bibfield  {author} {\bibinfo {author} {\bibfnamefont {Y.}~\bibnamefont
  {Fujita}}, \bibinfo {author} {\bibfnamefont {H.}~\bibnamefont {Fujita}},
  \bibinfo {author} {\bibfnamefont {T.}~\bibnamefont {Adachi}}, \bibinfo
  {author} {\bibfnamefont {C.~L.}\ \bibnamefont {Bai}}, \bibinfo {author}
  {\bibfnamefont {A.}~\bibnamefont {Algora}}, \bibinfo {author} {\bibfnamefont
  {G.~P.~A.}\ \bibnamefont {Berg}}, \bibinfo {author} {\bibfnamefont
  {P.}~\bibnamefont {von Brentano}}, \bibinfo {author} {\bibfnamefont
  {G.}~\bibnamefont {Col\`o}}, \bibinfo {author} {\bibfnamefont
  {M.}~\bibnamefont {Csatl\'os}}, \bibinfo {author} {\bibfnamefont {J.~M.}\
  \bibnamefont {Deaven}}, \bibinfo {author} {\bibfnamefont {E.}~\bibnamefont
  {Estevez-Aguado}}, \bibinfo {author} {\bibfnamefont {C.}~\bibnamefont
  {Fransen}}, \bibinfo {author} {\bibfnamefont {D.}~\bibnamefont {De~Frenne}},
  \bibinfo {author} {\bibfnamefont {K.}~\bibnamefont {Fujita}}, \bibinfo
  {author} {\bibfnamefont {E.}~\bibnamefont {Ganio\ifmmode~\breve{g}\else
  \u{g}\fi{}lu}}, \bibinfo {author} {\bibfnamefont {C.~J.}\ \bibnamefont
  {Guess}}, \bibinfo {author} {\bibfnamefont {J.}~\bibnamefont {Guly\'as}},
  \bibinfo {author} {\bibfnamefont {K.}~\bibnamefont {Hatanaka}}, \bibinfo
  {author} {\bibfnamefont {K.}~\bibnamefont {Hirota}}, \bibinfo {author}
  {\bibfnamefont {M.}~\bibnamefont {Honma}}, \bibinfo {author} {\bibfnamefont
  {D.}~\bibnamefont {Ishikawa}}, \bibinfo {author} {\bibfnamefont
  {E.}~\bibnamefont {Jacobs}}, \bibinfo {author} {\bibfnamefont
  {A.}~\bibnamefont {Krasznahorkay}}, \bibinfo {author} {\bibfnamefont
  {H.}~\bibnamefont {Matsubara}}, \bibinfo {author} {\bibfnamefont
  {K.}~\bibnamefont {Matsuyanagi}}, \bibinfo {author} {\bibfnamefont
  {R.}~\bibnamefont {Meharchand}}, \bibinfo {author} {\bibfnamefont
  {F.}~\bibnamefont {Molina}}, \bibinfo {author} {\bibfnamefont
  {K.}~\bibnamefont {Muto}}, \bibinfo {author} {\bibfnamefont {K.}~\bibnamefont
  {Nakanishi}}, \bibinfo {author} {\bibfnamefont {A.}~\bibnamefont {Negret}},
  \bibinfo {author} {\bibfnamefont {H.}~\bibnamefont {Okamura}}, \bibinfo
  {author} {\bibfnamefont {H.~J.}\ \bibnamefont {Ong}}, \bibinfo {author}
  {\bibfnamefont {T.}~\bibnamefont {Otsuka}}, \bibinfo {author} {\bibfnamefont
  {N.}~\bibnamefont {Pietralla}}, \bibinfo {author} {\bibfnamefont
  {G.}~\bibnamefont {Perdikakis}}, \bibinfo {author} {\bibfnamefont
  {L.}~\bibnamefont {Popescu}}, \bibinfo {author} {\bibfnamefont
  {B.}~\bibnamefont {Rubio}}, \bibinfo {author} {\bibfnamefont
  {H.}~\bibnamefont {Sagawa}}, \bibinfo {author} {\bibfnamefont
  {P.}~\bibnamefont {Sarriguren}}, \bibinfo {author} {\bibfnamefont
  {C.}~\bibnamefont {Scholl}}, \bibinfo {author} {\bibfnamefont
  {Y.}~\bibnamefont {Shimbara}}, \bibinfo {author} {\bibfnamefont
  {Y.}~\bibnamefont {Shimizu}}, \bibinfo {author} {\bibfnamefont
  {G.}~\bibnamefont {Susoy}}, \bibinfo {author} {\bibfnamefont
  {T.}~\bibnamefont {Suzuki}}, \bibinfo {author} {\bibfnamefont
  {Y.}~\bibnamefont {Tameshige}}, \bibinfo {author} {\bibfnamefont
  {A.}~\bibnamefont {Tamii}}, \bibinfo {author} {\bibfnamefont {J.~H.}\
  \bibnamefont {Thies}}, \bibinfo {author} {\bibfnamefont {M.}~\bibnamefont
  {Uchida}}, \bibinfo {author} {\bibfnamefont {T.}~\bibnamefont {Wakasa}},
  \bibinfo {author} {\bibfnamefont {M.}~\bibnamefont {Yosoi}}, \bibinfo
  {author} {\bibfnamefont {R.~G.~T.}\ \bibnamefont {Zegers}}, \bibinfo {author}
  {\bibfnamefont {K.~O.}\ \bibnamefont {Zell}}, \ and\ \bibinfo {author}
  {\bibfnamefont {J.}~\bibnamefont {Zenihiro}},\ }\href {\doibase
  10.1103/PhysRevLett.112.112502} {\bibfield  {journal} {\bibinfo  {journal}
  {Phys. Rev. Lett.}\ }\textbf {\bibinfo {volume} {112}},\ \bibinfo {pages}
  {112502} (\bibinfo {year} {2014})}\BibitemShut {NoStop}%
\bibitem [{\citenamefont {Fujita}\ \emph {et~al.}(2015)\citenamefont {Fujita},
  \citenamefont {Fujita}, \citenamefont {Adachi}, \citenamefont {Susoy},
  \citenamefont {Algora}, \citenamefont {Bai}, \citenamefont {Col\`o},
  \citenamefont {Csatl\'os}, \citenamefont {Deaven}, \citenamefont
  {Estevez-Aguado}, \citenamefont {Guess}, \citenamefont {Guly\'as},
  \citenamefont {Hatanaka}, \citenamefont {Hirota}, \citenamefont {Honma},
  \citenamefont {Ishikawa}, \citenamefont {Krasznahorkay}, \citenamefont
  {Matsubara}, \citenamefont {Meharchand}, \citenamefont {Molina},
  \citenamefont {Nakada}, \citenamefont {Okamura}, \citenamefont {Ong},
  \citenamefont {Otsuka}, \citenamefont {Perdikakis}, \citenamefont {Rubio},
  \citenamefont {Sagawa}, \citenamefont {Sarriguren}, \citenamefont {Scholl},
  \citenamefont {Shimbara}, \citenamefont {Stephenson}, \citenamefont {Suzuki},
  \citenamefont {Tamii}, \citenamefont {Thies}, \citenamefont {Yoshida},
  \citenamefont {Zegers},\ and\ \citenamefont {Zenihiro}}]{PhysRevC.91.064316}%
  \BibitemOpen
  \bibfield  {author} {\bibinfo {author} {\bibfnamefont {Y.}~\bibnamefont
  {Fujita}}, \bibinfo {author} {\bibfnamefont {H.}~\bibnamefont {Fujita}},
  \bibinfo {author} {\bibfnamefont {T.}~\bibnamefont {Adachi}}, \bibinfo
  {author} {\bibfnamefont {G.}~\bibnamefont {Susoy}}, \bibinfo {author}
  {\bibfnamefont {A.}~\bibnamefont {Algora}}, \bibinfo {author} {\bibfnamefont
  {C.~L.}\ \bibnamefont {Bai}}, \bibinfo {author} {\bibfnamefont
  {G.}~\bibnamefont {Col\`o}}, \bibinfo {author} {\bibfnamefont
  {M.}~\bibnamefont {Csatl\'os}}, \bibinfo {author} {\bibfnamefont {J.~M.}\
  \bibnamefont {Deaven}}, \bibinfo {author} {\bibfnamefont {E.}~\bibnamefont
  {Estevez-Aguado}}, \bibinfo {author} {\bibfnamefont {C.~J.}\ \bibnamefont
  {Guess}}, \bibinfo {author} {\bibfnamefont {J.}~\bibnamefont {Guly\'as}},
  \bibinfo {author} {\bibfnamefont {K.}~\bibnamefont {Hatanaka}}, \bibinfo
  {author} {\bibfnamefont {K.}~\bibnamefont {Hirota}}, \bibinfo {author}
  {\bibfnamefont {M.}~\bibnamefont {Honma}}, \bibinfo {author} {\bibfnamefont
  {D.}~\bibnamefont {Ishikawa}}, \bibinfo {author} {\bibfnamefont
  {A.}~\bibnamefont {Krasznahorkay}}, \bibinfo {author} {\bibfnamefont
  {H.}~\bibnamefont {Matsubara}}, \bibinfo {author} {\bibfnamefont
  {R.}~\bibnamefont {Meharchand}}, \bibinfo {author} {\bibfnamefont
  {F.}~\bibnamefont {Molina}}, \bibinfo {author} {\bibfnamefont
  {H.}~\bibnamefont {Nakada}}, \bibinfo {author} {\bibfnamefont
  {H.}~\bibnamefont {Okamura}}, \bibinfo {author} {\bibfnamefont {H.~J.}\
  \bibnamefont {Ong}}, \bibinfo {author} {\bibfnamefont {T.}~\bibnamefont
  {Otsuka}}, \bibinfo {author} {\bibfnamefont {G.}~\bibnamefont {Perdikakis}},
  \bibinfo {author} {\bibfnamefont {B.}~\bibnamefont {Rubio}}, \bibinfo
  {author} {\bibfnamefont {H.}~\bibnamefont {Sagawa}}, \bibinfo {author}
  {\bibfnamefont {P.}~\bibnamefont {Sarriguren}}, \bibinfo {author}
  {\bibfnamefont {C.}~\bibnamefont {Scholl}}, \bibinfo {author} {\bibfnamefont
  {Y.}~\bibnamefont {Shimbara}}, \bibinfo {author} {\bibfnamefont {E.~J.}\
  \bibnamefont {Stephenson}}, \bibinfo {author} {\bibfnamefont
  {T.}~\bibnamefont {Suzuki}}, \bibinfo {author} {\bibfnamefont
  {A.}~\bibnamefont {Tamii}}, \bibinfo {author} {\bibfnamefont {J.~H.}\
  \bibnamefont {Thies}}, \bibinfo {author} {\bibfnamefont {K.}~\bibnamefont
  {Yoshida}}, \bibinfo {author} {\bibfnamefont {R.~G.~T.}\ \bibnamefont
  {Zegers}}, \ and\ \bibinfo {author} {\bibfnamefont {J.}~\bibnamefont
  {Zenihiro}},\ }\href {\doibase 10.1103/PhysRevC.91.064316} {\bibfield
  {journal} {\bibinfo  {journal} {Phys. Rev. C}\ }\textbf {\bibinfo {volume}
  {91}},\ \bibinfo {pages} {064316} (\bibinfo {year} {2015})}\BibitemShut
  {NoStop}%
\bibitem [{\citenamefont {Goodman}(1984)}]{PhysRevC.29.1887}%
  \BibitemOpen
  \bibfield  {author} {\bibinfo {author} {\bibfnamefont {A.~L.}\ \bibnamefont
  {Goodman}},\ }\href {\doibase 10.1103/PhysRevC.29.1887} {\bibfield  {journal}
  {\bibinfo  {journal} {Phys. Rev. C}\ }\textbf {\bibinfo {volume} {29}},\
  \bibinfo {pages} {1887} (\bibinfo {year} {1984})}\BibitemShut {NoStop}%
\bibitem [{\citenamefont {Alhassid}\ \emph {et~al.}(1988)\citenamefont
  {Alhassid}, \citenamefont {Bush},\ and\ \citenamefont
  {Levit}}]{PhysRevLett.61.1926}%
  \BibitemOpen
  \bibfield  {author} {\bibinfo {author} {\bibfnamefont {Y.}~\bibnamefont
  {Alhassid}}, \bibinfo {author} {\bibfnamefont {B.}~\bibnamefont {Bush}}, \
  and\ \bibinfo {author} {\bibfnamefont {S.}~\bibnamefont {Levit}},\ }\href
  {\doibase 10.1103/PhysRevLett.61.1926} {\bibfield  {journal} {\bibinfo
  {journal} {Phys. Rev. Lett.}\ }\textbf {\bibinfo {volume} {61}},\ \bibinfo
  {pages} {1926} (\bibinfo {year} {1988})}\BibitemShut {NoStop}%
\bibitem [{\citenamefont {Dang}(2007)}]{PhysRevC.76.064320}%
  \BibitemOpen
  \bibfield  {author} {\bibinfo {author} {\bibfnamefont {N.~D.}\ \bibnamefont
  {Dang}},\ }\href {\doibase 10.1103/PhysRevC.76.064320} {\bibfield  {journal}
  {\bibinfo  {journal} {Phys. Rev. C}\ }\textbf {\bibinfo {volume} {76}},\
  \bibinfo {pages} {064320} (\bibinfo {year} {2007})}\BibitemShut {NoStop}%
\bibitem [{\citenamefont {Dang}\ \emph {et~al.}(1993)\citenamefont {Dang},
  \citenamefont {Ring},\ and\ \citenamefont {Rossignoli}}]{PhysRevC.47.606}%
  \BibitemOpen
  \bibfield  {author} {\bibinfo {author} {\bibfnamefont {N.~D.}\ \bibnamefont
  {Dang}}, \bibinfo {author} {\bibfnamefont {P.}~\bibnamefont {Ring}}, \ and\
  \bibinfo {author} {\bibfnamefont {R.}~\bibnamefont {Rossignoli}},\ }\href
  {\doibase 10.1103/PhysRevC.47.606} {\bibfield  {journal} {\bibinfo  {journal}
  {Phys. Rev. C}\ }\textbf {\bibinfo {volume} {47}},\ \bibinfo {pages} {606}
  (\bibinfo {year} {1993})}\BibitemShut {NoStop}%
\bibitem [{\citenamefont {Gambacurta}\ \emph {et~al.}(2013)\citenamefont
  {Gambacurta}, \citenamefont {Lacroix},\ and\ \citenamefont
  {Sandulescu}}]{PhysRevC.88.034324}%
  \BibitemOpen
  \bibfield  {author} {\bibinfo {author} {\bibfnamefont {D.}~\bibnamefont
  {Gambacurta}}, \bibinfo {author} {\bibfnamefont {D.}~\bibnamefont {Lacroix}},
  \ and\ \bibinfo {author} {\bibfnamefont {N.}~\bibnamefont {Sandulescu}},\
  }\href {\doibase 10.1103/PhysRevC.88.034324} {\bibfield  {journal} {\bibinfo
  {journal} {Phys. Rev. C}\ }\textbf {\bibinfo {volume} {88}},\ \bibinfo
  {pages} {034324} (\bibinfo {year} {2013})}\BibitemShut {NoStop}%
\end{thebibliography}%

\end{document}